\documentclass[preprint]{aastex}

\newcommand{\unit}[1]{\ensuremath{\, \mathrm{#1}}}

\usepackage{amsmath}
\usepackage[]{hyperref}

\shorttitle{The Chirpolator \& The Chimageator}
\shortauthors{Bannister \& Cornwell}

\begin{document}

\title{Two efficient, new techniques for detecting dispersed radio pulses with interferometers:\\
The Chirpolator and The Chimageator}

\author{K.~W.~Bannister\altaffilmark{1,2} and T.~J.~Cornwell}
\affil{CSIRO Astronomy and Space Sciences}
\email{k.bannister@physics.usyd.edu.au}

\altaffiltext{1}{Sydney Institute for Astronomy (SIfA), School of Physics, The University of Sydney}
\altaffiltext{2}{ARC Centre of Excellence for All-sky Astrophysics (CAASTRO) }

\begin{abstract}
Searching for dispersed radio pulses in interferometric data is of great scientific interest, but poses a formidable computational burden. Here we present two efficient, new antenna-coherent solutions: The Chirpolator and The Chimageator. We describe the equations governing both techniques and propose a number of novel optimisations. We compare the implementation costs of our techniques with classical methods using three criteria:  the operations rates (1) before and (2) after the integrate-and-dump stage, and  (3) the data rate directly after the integrate-and-dump stage. When compared with classical methods, our techniques excel in the regime of sparse arrays, where they both require substantially lower data rates, and The Chirpolator requires a much lower post-integrator operations rate. In general, our techniques require more pre-integrator operations than the classical ones. We argue that the data and operations rates required by our techniques are better matched to future supercomputer architectures, where the arithmetic capability is outstripping the bandwidth capability. Our techniques are, therefore, viable candidates for deploying on future interferometers such as the Square Kilometer Array.
\end{abstract}


\section{Introduction}

\subsection{Scientific Motivation}
Studying the high time resolution radio sky has illuminated the physics of our Galaxy, enabled exquisite measurements of physics at the extremes of gravity, density and magnetic field and uncovered a plethora of exotic objects. The main class of object enabling these measurements is the radio pulsar: a rapidly rotating magnetic neutron star which emits periodic, short pulses at radio frequencies.  While pulsars are interesting astrophysical laboratories in their own right, they can also be used to test predictions of General Relativity through observations of single pulsar systems (e.g. \citeauthor{Kramer06gr}, \citeyear{Kramer06gr}), search for gravitational waves with groups of pulsars (\citeauthor{Yardley10}, \citeyear{Yardley10} and references therein), and test theories of matter at the most extreme densities (e.g. \citeauthor{Demorest10}, \citeyear{Demorest10}). The short pulses emitted by pulsars undergo propagation effects during their passage through the Galactic Interstellar Medium (ISM), which enables measurements of the Galactic  magnetic field structure (e.g. \citeauthor{Vaneck11} \citeyear{Vaneck11}) and free electron density (e.g. \citeauthor{Cordes02} \citeyear{Cordes02}). There are also many more pulsars to be found: barely 2000 of the estimated 30000 \citep{Lorimer06} have been detected.

Radio pulsars are chiefly discovered by searching for periodic, dispersed radio emission. By contrast, searches for single pulses of radio emission have uncovered other types of objects, of which the most widely accepted are the so-called Rotating Radio Transients (RRATs) \citep{mclaughlin2006trb}. RRATs, like pulsars, are rotating neutron stars but emit only sporadically, and are being studied because they may hold the key to the so-called `missing supernova problem' \citep{Keane08}. Searches for single pulses have also yielded a number of intriguing short-duration radio transients, which do not fit the classical models of pulsars or RRATs (e.g. \citeauthor{lorimer2007bmr}, \citeyear{lorimer2007bmr}, \citeauthor{BurkeSpolaor11lg}, \citeyear{BurkeSpolaor11lg}, \citeauthor{Keane11}, \citeyear{Keane11}).

In spite of these discoveries, there is still much to do, as the the parameter space of radio transients is relatively poorly explored \citep{cordes2004drs}. Exploring this parameter space opens the potential for  discovering new objects and physics. These motivations are behind at least eight ongoing pulsar and single-pulse surveys \citep{Mclauglin11}, and more surveys are in the late stages of planning (e.g. \citeauthor{Macquart10}, \citeyear{Macquart10}).

\subsection{Improving pulsar and single-pulse surveys}

When surveying for pulsars and single-pulse sources, a desirable figure of merit is the product of instantaneous sensitivity and field of view, known as `survey speed'. Improving survey speed has three important consequences: (1) it reduces the integration time required to reach a flux density limit for periodic sources; (2) it reduces the computational requirements to search for pulsars in tight binary systems\footnote{Pulsars in very small orbits are the provide the most powerful tests of General Relativity.}; (3) it enables a deeper search of the parameter space for single-pulse transients, in terms of rarity or faintness.

In recent years, the optimal approach for maximizing survey speed has been to use large steerable and in-earth single dish telescopes feeding multi-beam receivers and wide-band electronics. This approach, however, is probably nearing its limit. Steerable single dish engineering has reached its practical limit at diameters of about 100~m and there are limited sites available for large in-earth reflectors, which also suffer from a limited view of the sky. Similarly, modern digital electronics have improved to the state where 2~GHz bandwidth is readily achievable, but the physics of pulsar emission and the interstellar medium limit the majority of the pulsar energy to the range 0.1 to 10~GHz, so an increase in processing bandwidth is unable to produce large improvements in sensitivity. Finally, the field of view of multi-beam receivers cannot grow indefinitely, as a large multi-beam receiver simply blocks too much of the dish aperture to be efficient.

The likely way forward for improvements in survey speed, therefore, is to use arrays of antennas. Using a large number of small antennas achieves simultaneously large field of view and sensitivity, and therefore survey speed. Using an array does have a dramatic cost, however: for configurations of interest to future surveys, the computational requirements increase to the point where the data processing becomes almost infeasible \citep{Smits09}.

\subsection{The problem: processing requirements}
The desire to use arrays of antennas to simultaneously obtain the large field of view and high sensitivity presents major data processing challenges, both in terms of the required data and operations rates. These  challenges have inspired a number of novel approaches. For example, \citet{Daishido00} proposed the so called Fast Fourier Transform Telescope (FFTT) for a pulsar survey, which used FFT beamforming, first proposed by \citet{Williams68}, and a square array geometry. This approach, also known as the Direct Imaging, has attractive properties in terms of operations rates and has been extended, with particular emphasis on 21~cm tomography, to arrays of regular, arbitrary hierarchies of grids by \citet{Tegmark10} and arbitrary array geometries by \citet{Morales08}. In a novel experiment, \citet{Janssen09} enhanced the field of view over standard techniques, by employing a uniform linear geometry and phased array beamforming, which introduced deliberate ambiguities into the synthesized beam (i.e. grating lobes). Recently, \citet{Trott11} proposed a method for searching for transient sources directly visibility space, rather than image-space. The visibility space approach is promising for arrays with sparse, arbitrary geometries, as the number of visibilities is much smaller than the number of pixels. But, is not yet clear whether this approach will achieve substantial computational savings.

Some of the above approaches rely on having control over the array geometry. In many cases, controlling the geometry is not possible because it is driven by other requirements e.g. the $uv$-coverage. The Square Kilometer Array (SKA) and its pathfinders fall into this category. In such cases, one can reduce the processing requirements by falling back to the reduced sensitivity of incoherent processing,  analyzing a smaller field of view than available from the primary beam \citep{Daddario10} or pointing all antennas at different parts of the sky in a ``Fly's Eye'' mode \citep{Macquart11}. Nonetheless, a fully-coherent, wide-field and computationally tractable system is a desirable goal.

We also note that the processing requirements are not limited to the number of arithmetic operations. In fact, in modern supercomputing problems, the processing bottleneck is not the number of arithmetic operations, but rather the data bandwidth into processor \citep{Leback08}.  The bandwidth bottleneck has been identified as a key problem for correlators for large interferometers, with a number of proposed solutions (e.g.  \citeauthor{Lutomirski11}, \citeyear{Lutomirski11}, \citeauthor{Carlson10}, \citeyear{Carlson10}). To our knowledge, the data requirements of classical fast transient detection techniques has not been discussed explicitly in the literature, so we consider them in this paper.

\subsection{Two new techniques}
It is in this context that we propose two new techniques, which we have named after the  frequency swept signals on which they operate (`chirps'). The first technique, which we call `The Chirpolator'\footnote{A portmanteau of `chirp' and `correlator'.} operates by correlating the chirps received by pairs of antennas. The second technique, which we call `The Chimageator'\footnote{A portmanteau of `chirp',  `image' and `correlator'. The `imaging' part is inspired by the Direct Imaging approach of \citet{Daishido00}.} operates by gridding the cross-multiplied voltages from all telescopes to form an image at every sampling time. Both techniques are applicable to arbitrary array configurations, exploit the full sensitivity of the telescope, and have substantially lower data rate requirements than classical coherent techniques. Thus, these new techniques may be favored over classical techniques in many regimes of computer economics and array geometry.

This paper is organized as follows: in section 2 we provide background of the problem of searching for dispersion emission in interferometric data, and the classical solutions. In section 3 we describe The Chirpolator and section 4 we describe The Chimageator. We describe a simple model to compare our techniques with classical results in section 5, and the results of this comparison in section 6. We discuss the implications of these two algorithms on future telescope design and science outcomes in section 7 and draw our conclusions in section 8. In the appendices we present detailed analysis of the algorithms, a number of novel optimizations, and a discussion of implementation considerations

\section{Background}

\subsection{Dispersion in the Interstellar Medium}
\label{sec:dispersion}

Before being received by telescopes on Earth, electromagnetic waves from an astronomical source must pass through the interstellar medium (ISM), a plasma containing non-relativistic unbound electrons. As it travels through the ISM, the wave undergoes dispersion, or frequency-dependent delay between two frequencies  $\nu_1$ and $\nu_2$ according to the following formula:

\begin{equation}
t  = \frac{e^2}{2 \pi m_e c} \mathrm{DM} \left ( \nu_2^{-2} - \nu_1^{-2} \right ) \label{eq:dmdelay}
\end{equation}

\noindent where $t$ is the time from the beginning of the pulse, $\nu_1 > \nu_2$,  and the physical constant is:

\begin{equation}
\mu = \frac{e^2}{2 \pi m_e c} \simeq 4.15 \mathrm{ms}
\end{equation}
\noindent for frequencies in GHz. The dispersion measure (DM) describes the number of electrons between the observer and the emitting source, defined as:

\begin{equation}
DM = \int_{0}^{d} n_e dl
\end{equation}

\noindent where $n_e$ is the electron density and $d$ is the distance to the source. DM is usually quoted in units of $\unit{cm^{-3} pc}$.

If a narrow pulse is emitted by a source, the frequency of the signal received at Earth will be exactly the form of the dispersion. For our analysis, we will consider the form of this dispersed pulse as a complex voltage time series. To form this time series, we will compute the instantaneous phase of the signal, by integrating the instantaneous frequency of the signal. The instantaneous frequency can be found by rearranging  Equation \ref{eq:dmdelay}, and we approximate it with a Taylor series around $t = T/2$, which yields:

\begin{eqnarray}
\nu_2(t) & = & \left (\nu_1^{-2} + \frac{t}{\mu DM} \right )^{-1/2} \label{eq:freqvst} \\
& = & \sum_{i=0}^{\infty} a_i (t - T/2)^i \label{eq:freqvst2}
\end{eqnarray}

\noindent where $\nu_2(t)$ is the instantaneous frequency of the signal, $t$ is the time from the beginning of the pulse, and $T$ is the time taken for the pulse to traverse the bandwidth of interest $B=\nu_1 - \nu_2$. The first three Taylor coefficients are:

\begin{eqnarray}
a_0 & = & \alpha^{-1/2} \\
a_1 & = & \frac{-1}{2} \frac{1}{\mu DM} \alpha^{-3/2} \\
a_2 & = & \frac{3}{8} \frac{1}{(\mu DM)^2}  \alpha^{-5/2}
\end{eqnarray}

\noindent where,

\begin{eqnarray}
\alpha & = & \nu_1^{-2} + \frac{T}{2 \mu DM} \\
& = & \frac{1}{2} \left ( (\nu_1 - B)^{-2} + \nu_1^{-2} \right )
\end{eqnarray}.

\begin{figure}
\centering
\includegraphics[width= \linewidth]{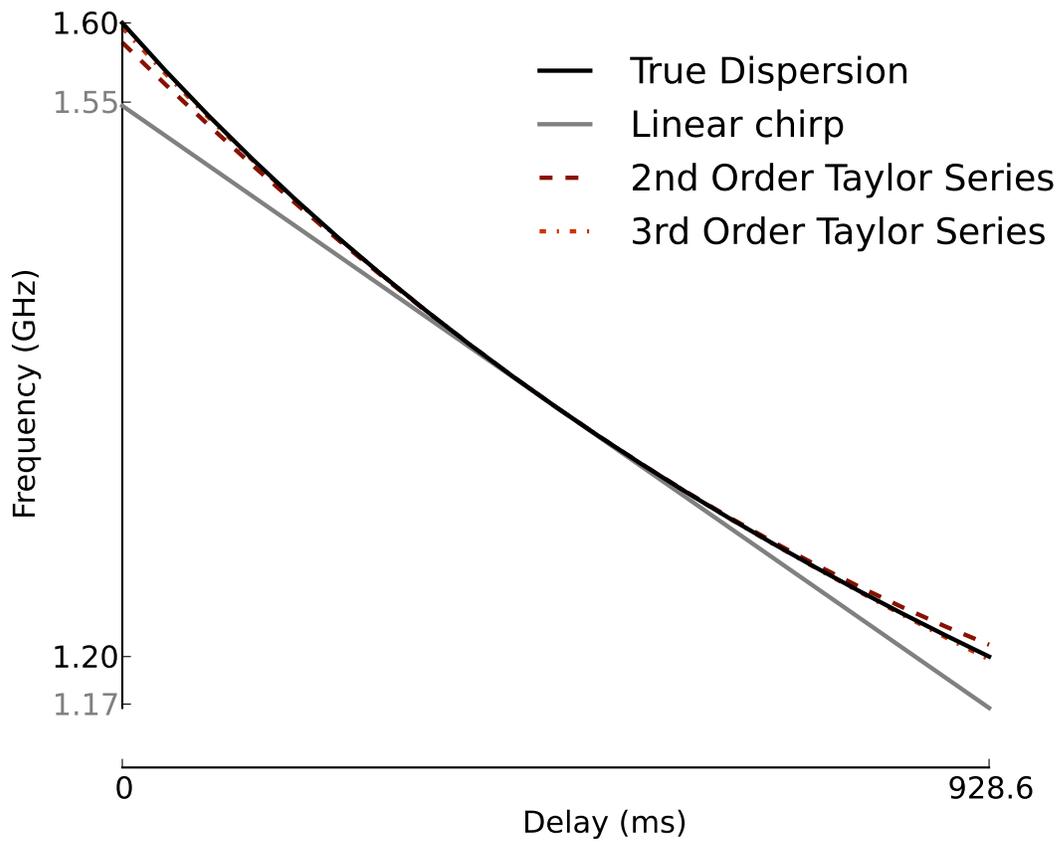}
\caption{Delay vs frequency for a pulse with DM of  $100 \unit{cm^{-3} pc}$, over a frequency range of 400~MHz centered at 1.4~GHz. The true dispersion (Equation \protect \ref{eq:dmdelay}) and the first (linear chirp), second and third order Taylor series approximations around the delay midpoint are shown. Third order approximation is barely visible as it very closely matches the true dispersion. }
\label{fig:delayvsfreq}
\end{figure}

A plot of the true dispersion law, and the linear and second order approximations is shown in Figure \ref{fig:delayvsfreq}.

To obtain the formula for the voltage time series of the signal received on Earth, we first write the phase of the signal, as given by:

\begin{eqnarray}
\phi(t) & = & 2 \pi \int_0^{t}{\nu_2(t') dt'} \\
& = & 4 \pi \mu DM \left (\nu_1^{-2} + \frac{t}{\mu DM} \right )^{1/2} \\
& = & 2 \pi \sum_{i=0}^{\infty}{\frac{a_i}{i+1}(t - T/2)^{i+1}} \\
& \simeq & 2 \pi \left [ \frac{T}{2}(-a_0 + a_1/2)  + t (a_0 - a_1 T/4) + t^2 a_1/2 \right] \label{eq:phit}
\end{eqnarray}

\noindent where we have expanded out the phase to the $i=1$ Taylor term. Finally, we can write the complex voltage as:

\begin{equation}
s(t)   =   \exp(j \phi(t)).
\end{equation}

In a typical telescope system, the absolute phase of the voltage (i.e. the constant term in Equation \ref{eq:phit}) is not important, and the fixed frequency ($t$ term) is removed by down-conversion, so only the $t^2$ and higher terms are relevant. Therefore, in the main text we approximate the dispersion with a signal for the form:

\begin{equation}
s(t)  =  \exp(\pi j \dot{f} t^2)
\end{equation}

\noindent where $\dot{f} = a_1 \simeq -B/T$ is the gradient of the linear frequency trajectory as shown in Figure \ref{fig:delayvsfreq}. Signals of this form are known as complex linear chirps. In the appendices we consider the higher order terms.

\subsection{A taxonomy of methods}

\begin{figure*}
\centering
\includegraphics[width=\textwidth]{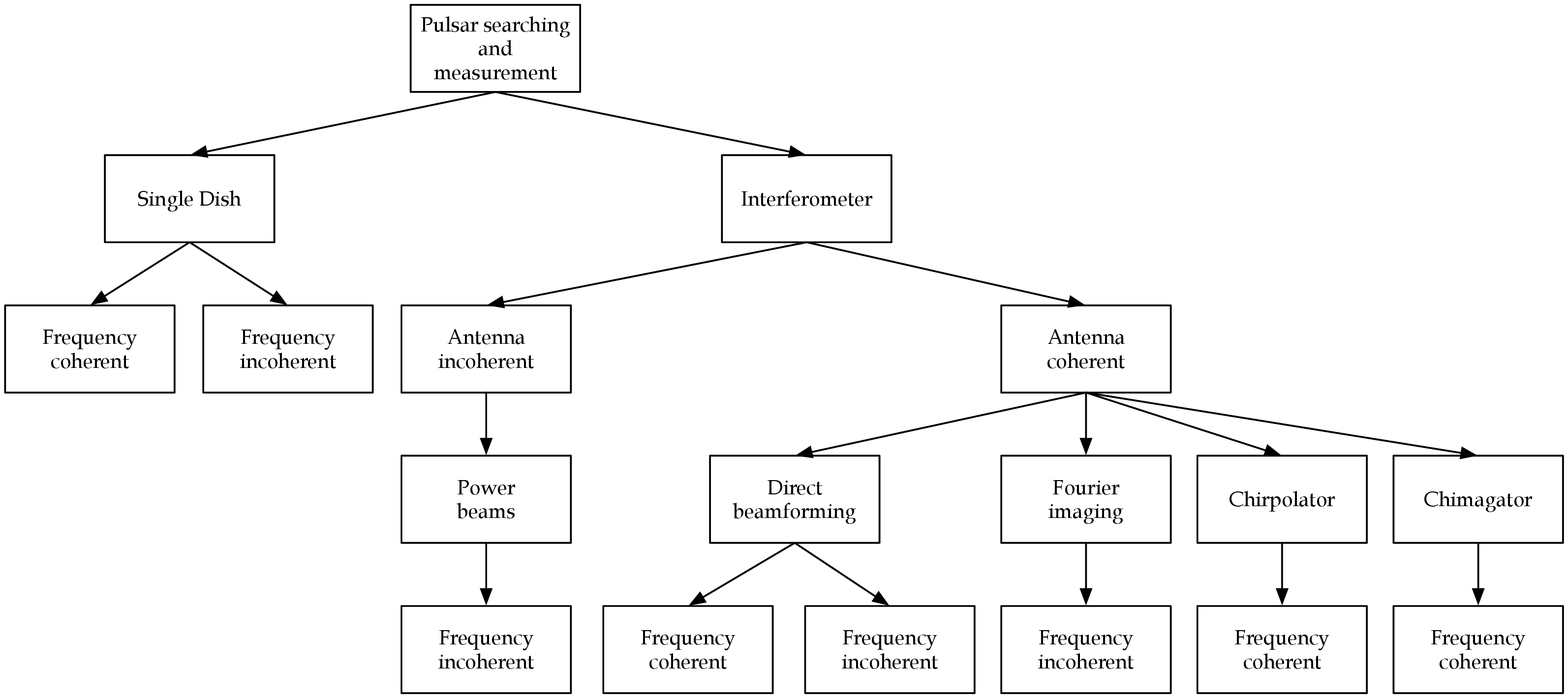}
\caption{Taxonomy of approaches to pulsar searching and measurement.}
\label{fig:taxonomy}
\end{figure*}

An astronomer wishing to search for, or study short duration radio pulses, may employ any one of a wide range of dedispersion and array processing techniques, as shown in Fig. \ref{fig:taxonomy}. General properties of these methods are shown in Table \ref{tab:methods}. We give more detailed descriptions of the classical methods in the following sections.

\begin{table*}[htdp]
\caption{Methods of searching for pulsars and scaling relations for various figures of merit. The columns are: the name of the method, the resolution in radians, the number of pixels in an image, and the scaling relation for the sensitivity with number of antennas $M$. Parabolic dishes and equal maximum baselines in the $u$ and $v$ directions have been assumed.}
\label{tab:method}
\begin{center}
\begin{tabular}{l c c c}\hline
Method & $\Delta \theta$ & $N_{\rm pix}$ & Sensitivity Scaling \\
& (rad) & \\
\hline
\\
Power beams & $1.17 \frac{\lambda}{D}$ (primary beam) & 1 & $\propto M^{1/2}$ \\
\\
Fourier imaging / Direct beamforming & $\frac{\lambda D}{b_{max}}$ & $\left ( \frac{b_{max}}{D} \right ) ^2$ & $\propto M $\\
\\
Chirpolator / Chimageator & $0.844 \frac{c}{B b_{\rm max}}$ &  $1.92 \left ( \frac{  B b_{\rm max} \lambda}{c D} \right )^2$ & $\propto M $ \\
\\
\hline
\end{tabular}
\end{center}
\label{tab:methods}
\end{table*}

\subsection{Dedispersion}

The effect of dispersion on a short-duration pulse is to smear it out in time. In order to determine the emitted pulse shape, or to detect the pulse with maximum signal to noise ratio, the effect of the dispersion must be undone, in a process known as dedispersion. Two methods of dedispersion can be employed, as described below.

\subsubsection{Incoherent dedispersion}
\label{sec:incoherent_dedispersion}

Incoherent dedispersion involves two steps. First, the raw telescope voltages are passed through an analysis filterbank, such as an analog filterbank, Fast Fourier Transform (FFT) or polyphase filterbank to form a set of channelized outputs. Each filter output is then squared, and integrated over an interval (typically 0.1-10ms) to form a spectrogram, or time-frequency plane. In the second step, a range of trial DMs are searched, by summing across frequency channels, after delaying each frequency channel according to the trial DM of interest. We call this method `frequency incoherent' because only the filterbank amplitudes are summed, and the phase information is discarded. Incoherent dedispersion is typically used in pulsar surveys because the filtering can be computed only once and a range of DM trials can be performed on the same filtered output  relatively cheaply.

While early workers used analog filterbanks, more recent projects digitally sample baseband voltage signals and perform the digital filtering and dedispersion in hardware and software.  \citet{taylor1974ddisp} proposed a computationally efficient method of forming performing incoherent dedispersion, known as the `tree' method, which requires fewer additions than a naive implementation, but assumes linear dispersion. The linear assumption can be relaxed by adding padding channels, which marginally increases the computational cost.

\subsubsection{Coherent dedispersion}

Coherent dedispersion operates on raw telescope voltages, and involves convolving the telescope voltages with the impulse response corresponding to the inverse of the ISM  (i.e. the inverse of Eq. \ref{eq:dmdelay}),  thereby forming the maximum signal-to-noise ratio filter, or `matched' filter. We call this method `frequency coherent' as the data are processed without discarding the phase. Coherent dedispersion is used during pulsar monitoring, when the DM is approximately known, as the inverse filtering preserves the emitted pulse shape more faithfully than incoherent dedispersion. Coherent dedispersion is not used for pulsar surveys, as the computational cost of performing multiple DM trials is prohibitive. Coherent dedispersion is most often performed on digitally sampled complex baseband signals.

\subsection{Array processing}

When using an array, the question arises of how to best combine the signal from two or more antennas. In this section we describe three common approaches.

\subsubsection{Power beams}
Power beams are formed by envelope detecting the output of each antenna, and summing the resulting powers across antennas. We call this method `antenna incoherent' as the envelope detection removes the phase information before the sum across antennas. The power beam is sensitive to the entire sky, as long as the integration time of the envelope detector is longer than the largest geometric delay, and is usually limited by the primary beam of the telescope antennas. The penalty for power beams is that the sensitivity is poor, as it scales as $M^{1/2}$, where M is the number of antennas. The output of the power beams can only be incoherently dedispersed, as the phase information is discarded by the envelope detector at the antenna.

\subsubsection{Direct beamforming (Tied array beams)}
\label{sec:direct_beamforming}
Direct beamforming involves delaying the voltage signal from each telescope to compensate for the array geometry and summing the resulting voltages. This technique is also known as `tied array beam forming'. This method `antenna coherent' as the phase information is preserved. The resulting beam  has the size of a synthesized beam which is much smaller than the telescope primary beam. Unlike power beams, the full array sensitivity is preserved as it scales with $M$. As a tied array beam provides a voltage stream, either coherent or incoherent dedispersion can be used. Multiple tied array beams can be deployed to increase the field of view.

\subsubsection{Fourier imaging}
\label{sec:fourier_imaging}
Fourier imaging involves cross-correlating the telescope voltages with one another to form a set of complex `visibilities' which are Fourier transformed to form an image. Cross correlation can be performed either by an initial filtering step followed by cross-multiplication (so called FX correlation), or cross correlation followed by a Fourier transform (so called XF correlation). Each pixel of the image must be separately incoherently dedispersed, as the pixels are spatially independent. Coherent dedispersion cannot be used because each pixel contains only amplitude information.

Fourier imaging achieves the full array sensitivity over the full primary beam of the individual antennas. As only incoherent dedispersion can be used, Fourier imaging is most suited to  surveys. Fourier imaging requires a  so called `corner-turn', or matrix transpose between the imaging and dedispersion stages, which can result in very high data rates between the two steps.  Fourier imaging has recently been used by \citet{Law11} and \cite{Wayth11} to giant pulses from the Crab pulsar.

\section{The Chirpolator}
\label{sec:chirpolator}

In this section we provide an intuitive description of The Chirpolator, and provide a derivation of the equations  beginning with the simplest two antenna case. We then extend the results to multiple antennas in 1D. Extensions to 3D telescope geometries, non-linear dispersion delay and novel techniques for efficiently implementing The Chirpolator are described in Appendix \ref{sec:app_chirpolator}.

\subsection{Intuitive Description}
\label{sec:chirp_intuitive}

Here we describe an overview of The Chirpolator to aid the intuition of the reader. Put simply, The Chirpolator exploits the observation that when a linear chirp received by one antenna, is multiplied by a delayed linear chirp received at another antenna, the result is a fixed-frequency tone whose frequency is proportional to the geometric delay. The DFT of these tones can be coherently combined across all antenna pairs to form a detection metric.

A more rigorous mathematical description is described in Section \ref{sec:two_antennas} and following.

\begin{enumerate}
\item We model the dispersed pulse from an astronomical source as a finite-duration linear chirp ($s(t)$); i.e. a signal whose frequency sweeps linearly across the bandwidth ($B$) in a time $T$ (see Figure \ref{fig:method}, top panel). Such a signal has a constant frequency gradient of $\dot{f} = B/T$.

\item This signal is received by two antennas, indexed $p$ and $q$. The signal  ($s_p(t)$) at antenna $p$i s delayed with respect to the signal ($s_q(t)$) at antenna $q$  by an unknown geometric delay ($\tau$).

\item The difference in frequency between the two signals is constant for the duration of the pulse, and is equal to $\dot{f} \tau$ (Figure \ref{fig:method}, top panel).

\item If we multiply the signal from antenna $p$ by the conjugate of the signal from antenna $q$, the result ($x_{pq}(t)$) is a tone at fixed frequency (Figure \ref{fig:method}, bottom-left panel). This multiplication is equivalent to `downconversion'  (also known as `mixing'), which shifts the frequency of a signal in a radio frequency system. In the mixing case, an incoming signal is multiplied by a fixed-frequency Local Oscillator (LO), and the result has a center frequency which is the \emph{difference} between the center frequency of the incoming signal, and the LO frequency. In our case, both the `LO' and the incoming signal are sweeping at the same rate ($\dot{f}$) but the frequency difference remains fixed. Thus, the signal at antenna $p$ is effectively `downconverted' by an `LO' (provided by antenna $q$) which is perfectly matched in frequency, yielding a fixed-frequency tone.

\item We have a fixed frequency tone, with unknown frequency ($\dot{f} \tau$) and duration $T$. In practise, this tone will also be contaminated by noise. By taking the Discrete Fourier Transform (DFT) of this signal ($X_{pq}[k]$), all the energy of the sinusoid is coherently added into a small number of DFT  bins, while the noise adds incoherently. Therefore, taking the DFT increases the signal-to-noise ratio by approximately the square root of the DFT length. Also, for most arrays of interest, the signal can be more compactly expressed in a DFT as the range of possible frequencies is much smaller than the number of samples (see Section \ref{sec:only_dft_baselines}), which reduces the downstream data and processing rates.

\item We repeat the above two steps for each antenna pair, and produce a DFT spectrum for each (Figure \ref{fig:method}, bottom-right panel). The spectrum has a peak at frequency $k_0$. This frequency is proportional to the geometric delay ($\tau$), which is in turn proportional to the baseline length ($b_{pq}$) and angle of arrival ($\theta$) (see Figure \ref{fig:freqvsbaseline}). The value of the peak of the spectrum ($X_{pq}[k_0]$) is a complex number  whose phase ($\Phi_{pq}$) is also a function of the geometric delay ($\tau$).

\item Finally we form an image, which is a detection metric ($P(\theta)$) for a range of trial directions of interest. To produce the detection metric for a given direction of interest, we compute the expected arrival frequency ($k_0'$) and DFT phase correction ($\Phi^*_{pq}$) for a given antenna pair. We then pick out the DFT bin at the expected frequency ($X_{pq}[k_0']$) and multiply by the phase correction ($\Phi^*_{pq}$) so that the bins for all pairs have the same absolute phase (See Figure \ref{fig:calerrors} ). A vector sum of the phase-corrected DFT bins over all antenna pairs is a coherent sum across all antennas, and yields a detection metric in the direction of interest.

\item In practise, both the time of arrival, and actual DM (equivalent to $T$) are not known in advance. Therefore, we repeat the above procedure in a sliding window fashion, and assuming a range of DMs. This repetition can be efficiently implemented using a number novel of techniques (see Section \ref{sec:implementation_optimizations}).

\end{enumerate}

\subsection{Two-antenna case}
\label{sec:two_antennas}

\begin{figure*}
\centering
\includegraphics[width=\textwidth]{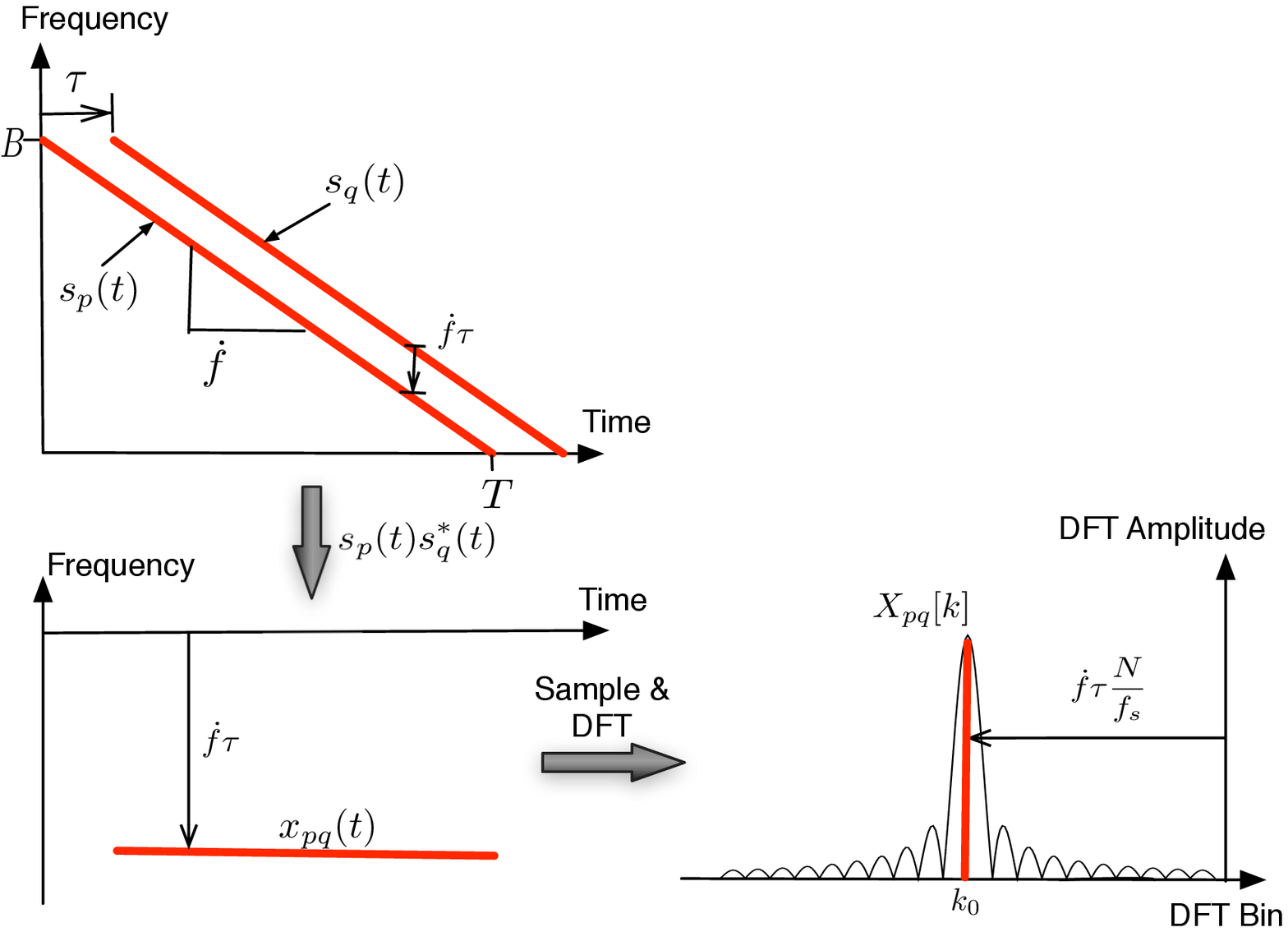}
\caption{Schematic illustrating the The Chirpolator operating with two antennas. Top left: Two linear chirps are received by antennas $p$ and $q$, with one delayed by $\tau$. Bottom left: After taking the product of the two voltage time series, the result $x_{pq}$ has constant frequency over most of the duration of the chirp. Bottom right: The DFT of $x_{pq}$ yields a peak at $k_0$.}
\label{fig:method}
\end{figure*}

In this section we develop a more rigorous description of The Chirpolator. We begin by considering a single pulse which has been dispersed by the ISM, which we approximate by a linear chirp impinging on an ideal (perfectly calibrated) two-antenna array. A schematic of the scheme is shown in Fig. \ref{fig:method}.

This technique was described by \citet{Gershman01} as the maximum likelihood detector for a single chirp, which they termed a `chirp beamformer'.

As described in Section \ref{sec:dispersion}, the voltage waveform received by an antenna can be written as a complex linear chirp with unit amplitude:
\begin{equation}
s(t) = \exp \left ( \pi j \dot{f} t^2 \right ) \label{eq:chirp}
\end{equation}
where $\dot{f} = -B/T$ is the chirp rate, B is the system bandwidth and T is the time taken for the chirp to cross the bandwidth. Assume this signal is received by two antennas, with the signal delayed at antenna $q$ by $\tau$ seconds with respect to the arrival at antenna $p$. The product of the chirp received by antenna $p$, with its delayed and conjugated counterpart from antenna $q$ is:

\begin{eqnarray}
x_{pq}(t) & =  & s_p(t) s_q^*(t) \label{eq:xpq} \\
& = & s(t) s^*(t - \tau) \\
& = & \exp \left ( \pi j \dot{f} t^2 \right ) \exp \left ( -\pi j \dot{f} (t - \tau)^2 \right )  \\
& = & \exp \left ( \pi j \dot{f}(2 t \tau - \tau^2) \right )
\end{eqnarray}

\noindent which is a complex sinusoid of frequency $\dot{f} \tau$ and phase $- \pi j \dot{f} \tau^2$. Taking the product in this way is also termed `mixing'. We have assumed here that $\tau \ll T$, which implies that the signals received by both antennas substantially overlap in time. If we have discrete-time sampling we simply replace $t \to n/f_s$ where $n$ is the sample number and $f_s$ is the sampling frequency. For complex Nyquist sampling $f_s = B$.

The sampled version of $x_{pq}(t)$ is, therefore

\begin{equation}
x_{pq}[n] = \exp \left ( \pi j \dot{f} \left (\frac{2 n \tau}{f_s} - \tau ^2 \right ) \right ) .
\end{equation}

If we take the Discrete Fourier Transform (DFT) of $x_{pq}[n]$ over N samples, where $N=f_s T = B T$, and by using the standard result of the DFT of a complex sinusoid of finite duration \footnote{Use the shift theorem, the Fourier transform of a delta function and the similarity theorem.}, we obtain:

\begin{eqnarray}
\label{eq:dft}
X_{pq}\left [k \right ] &   =  &  DFT\{x_{pq}[n] \}   \\
& = & \sum_{n=0}^{N-1}{\exp \left ( \frac{-2 \pi j n k }{N} \right ) \exp \left ( \pi j \dot{f} \left(\frac{2 n \tau}{f_s} - \tau^2 \right ) \right ) } \\
& = & \sum_{n=0}^{N-1}{\exp \left ( \frac{ -2 \pi j n k }{N}  + \frac{2 \pi j n\dot{f}  \tau}{f_s}  \right ) \exp \left ( -\pi j \dot{f} \tau^2 \right )  } \\
& = & \exp \left ( -\pi j \dot{f} \tau^2 \right ) \sum_{n=0}^{N-1}{\exp \left (\frac{-2 \pi j n }{N} \left ( k - \frac{\dot{f} \tau N }{f_s} \right )   \right ) } \\
& = & \Phi_{pq}(k - k_0) D_N(k - k_0), \label{eq:phidn}
\end{eqnarray}

where $k_0$ the frequency of $x_{pq}[k]$ (in units of DFT bins), given by

\begin{eqnarray}
k_0 & = & \frac{\dot{f} \tau N}{f_s} \\
& = & 	\frac{(B/T) \tau (f_s T)}{f_s} \\
& = & B \tau ,
\end{eqnarray}

\noindent $D_N(x)$ is a real-valued amplitude term, whose shape is the Dirichlet kernel, defined as:

\begin{equation}
D_N(x) = \left\{
\begin{array}{l l}
N & \quad x = 0\\
\frac{\sin( \pi x)}{\sin( \pi x /N )} & \quad x \neq 0, -N < x < N   \label{eq:dn}
\end{array} \right.
\end{equation}

\noindent and $\Phi_{pq}(x)$ is a unit-amplitude complex phase term given by:

\begin{eqnarray}
\Phi_{pq}(x) & = &  \exp \left ( -\pi j  \dot{f} \tau^2 \right ) \exp \left ( -\pi j  x \frac{N-1}{N} \right ) \\ \label{eq:phipq1}
& \simeq &  \exp \left ( -\pi j  x \right ). \label{eq:phipq}
\end{eqnarray}

For the two antenna case, we can write the geometric delay $\tau$ simply as:

\begin{equation}
\tau = \frac{b_{pq} \sin \theta}{c}
\end{equation}

\noindent where $b_{pq}$ is the distance between the antennas, $\theta$ is the angle of the source off the phase center, and $c$ is the speed of light in the medium. The discrete frequency of $x_{pq}$, $k_0$, corresponds to the position of the peak in the spectrum $X_{pq}[k]$, and is related to the baseline length, angle of arrival and bandwidth by:

\begin{eqnarray}
\label{eq:f0}
k_0 & = & B \tau \label{eq:f0Bt} \\
& = & B \frac{b_{pq} \sin \theta}{c}  \label{eq:f0Bbaseline}
\end{eqnarray}

\begin{figure}
\centering
\includegraphics[width= \linewidth]{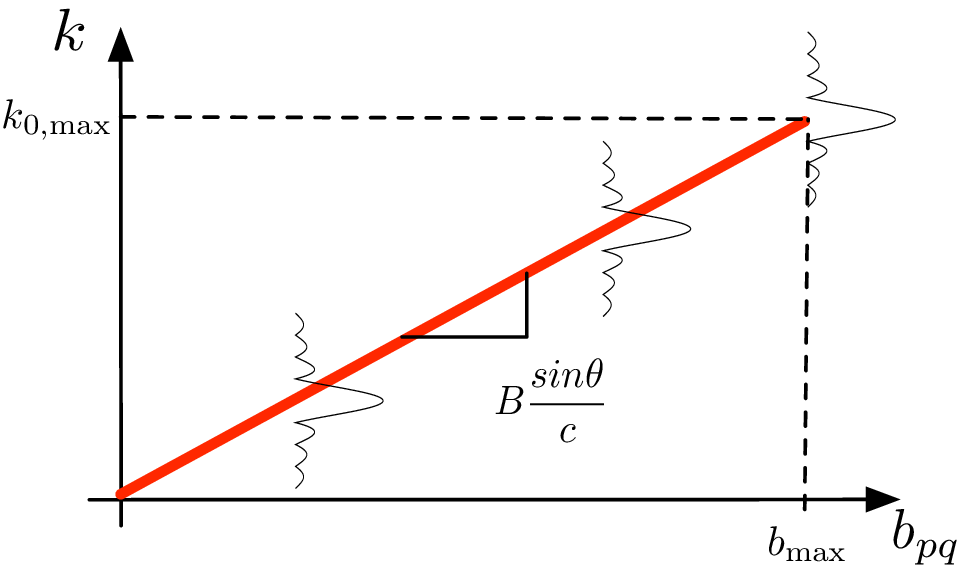}
\caption{Response of The Chirpolator to a source off the phase center. As the baseline length ($b_{pq}$) between antennas increases, the position of the peak in the DFT ($k_0$) increases linearly, with a gradient given by $B \sin(\theta)/c$ (see Equation \ref{eq:f0Bbaseline}). The amplitude of the DFT, $D_N(k - k_0)$, is shown to illustrate that there is some smearing of the signal around the expected frequency $k_0$.}
\label{fig:freqvsbaseline}
\end{figure}

Thus, the frequency of the mixed signal is linearly related to the $\sin$ of the angle arrival and the baseline length, as sketched in Figure \ref{fig:freqvsbaseline}.

To get an idea of the important factors in the above expressions, we can substitute typical values for current medium-sized dish-based radio telescopes. For a  baseline of 1~km, bandwidth of 400~MHz, dispersion delay of 1~s and beamwidth of 1 degree, $\tau^2 \ll 1$ and the first exponential term of Equation \ref{eq:phipq1} approaches $1$. In the same regime, the number of samples $N$ is large, of the order  $N = f_s T = B T > 10^5 $. The large number of samples has two consequences. Firstly,  in Equation \ref{eq:phipq1}, $(N-1)/N \to 1$, and, most importantly the amplitude term, $D_N(x)$ in Equation \ref{eq:dn} has only a small region of support around $k_0$. This allows computational savings by allowing us to truncate the computation of DFT bins to a few bins centered around $k_0$, meaning that calculation of the full DFT spectrum is not required, and downstream processing is also considerably reduced.

\subsection{Multiple telescopes in 1D}

To detect a chirp with a given $\dot{f}$ coming from an unknown direction, we form a detection metric, or intensity image, over a range of directions of interest. The detection metric is formed by phasing up results from all pairs of antennas. For simplicity we assume the array is perfectly calibrated, we ignore the smearing from the $D_N(x)$ term. Assuming the DFT spectrum is a single delta function, with all energy in the bin:

\begin{equation}
k_0' = \rm{round}(k_0). \label{eq:f0dash}
\end{equation}

\noindent then for a particular direction of interest $\theta$, we compute $k_0'$ for each antenna pair using Equations \ref{eq:f0Bbaseline} and  \ref{eq:f0dash}, and then compute the value of a single DFT bin $X_{pq}[k_0']$. We can essentially phase up each antenna pair by multiplying by the conjugate of the known phase term in Equation \ref{eq:phipq} and a detection metric can be formed by performing a vector sum across the phased-up antenna pairs according to:

\begin{equation}
\label{eq:pth}
P(\theta)  = \sum_{p=0}^{M-1} \sum_{q=p+1}^{M-1} \Phi^*_{pq}(k_0' - k)  X_{pq}[k_0']
\end{equation}

The procedure can be repeated for a range of $\theta$. If the direction of interest $\theta$ and the actual angle of arrival coincide, the DFT will have a peak at $k_0'$ with a value of $X_{pq}[k_0']$. By substituting Equation \ref{eq:phidn}, $P(\theta)$ reduces to:

\begin{eqnarray}
P(\theta)&  = &  \sum_{p=0}^{M-1} \sum_{q=p+1}^{M-1}  \Phi^*_{pq}( k_0' -k_0) \Phi_{pq}(k_0' - k_0) D_N(k_0' - k_0) \label{eq:pthsum} \\
& = &  \sum_{p=0}^{M-1} \sum_{q=p+1}^{M-1} D_N(k_0' - k_0) \\
& \simeq &  \sum_{p=0}^{M-1} \sum_{q=p+1}^{M-1} D_N(0) \\
& = & N\frac{M(M-1)}{2}
\end{eqnarray}

In general, the quantity $P(\theta)$ will be complex-valued. We are not interested in the absolute phase of the signal, so a more useful metric, for thresholding is:

\begin{equation}
\label{eq:eth}
E(\theta) = |P(\theta)|^2
\end{equation}

The fact the sum over antenna pairs is a vector sum means the resulting signal-to-noise scales with M rather than the $M^{1/2}$ scaling for non-coherent addition.

The time sequence of $E(\theta)$ for a particular value of $\theta$ can be considered a typical time sequence of power measurements, and can be subjected to the usual pulsar detection methods such as periodicity and acceleration searches.

\subsection{Compensating for the smearing in $D_N(x)$}
\label{sec:improving_snr}

Equations \ref{eq:f0dash}  and \ref{eq:pthsum} assume that all the energy is concentrated in a single bin. For an arbitrary angle of arrival, this is not the case, and in the worst case, the energy can be spread over all the bins in the DFT (see Figure \ref{fig:method}). For large $N$, $D_N(x)$ has relatively compact support, so we can truncate the number of DFT bins we compute, as well as the number of bins which need to be summed for a given direction of arrival and antenna pair. We can choose, therefore, to truncate the computations to $2F + 1$ bins centered around $k_0'$. In practice one can choose a value of $F$ that provides the best trade between computational cost and signal-to-noise ratio (SNR).

To capture the energy with support $[-F, F]$ around $k_0$, we perform a matched filter operation against the expected amplitude response function, which is the shifted $D_N(x)$. Therefore, equation \ref{eq:pthsum} can be trivially generalized to:

\begin{eqnarray}
P(\theta) & = &\sum_{p=0}^{M-1} \sum_{q=p+1}^{M-1} \sum_{k = -F}^{k = +F} \Phi^*_{pq}( k + k_0' - k_0 )  D_N(k - k_0) X_{pq}[k + k_0']
\end{eqnarray}

\section{The Chimageator}

We now describe an alternative method for combining signals from multiple telescopes based on gridding cross-multiplied voltages. Once again, we begin with an intuitive description and provide more mathematical rigour in later sections.

\subsection{Intuitive Description}

Here we describe an overview of The Chimageator to aid the intuition of the reader. The first three steps of The Chimageator are exactly the same as The Chirpolator (Section \ref{sec:chirp_intuitive}), i.e. The Chimageator  exploits the observation that when a linear chirp received by one antenna is multiplied by a delayed linear chirp received at another antenna, the result is a fixed-frequency tone whose frequency is proportional to the geometric delay. The difference between the two techniques is how the cross-multiplied data are combined: The Chimageator exploits an efficient spatial FFT at each sample time. The result is a dynamic spectrum where the chirp deposits energy along a linear trajectory in the time-spatial frequency plane. The gradient of the trajectory is proportional to the geometric delay. We sum DFT bins along a range of trial trajectories to form a detection metric.

\begin{enumerate}

\item We begin as with The Chirpolator, by assuming a linear chirp which sweeps  across the bandwidth ($B$) in time $T$, with gradient $\dot{f} = B/T$.

\item As with The Chirpolator, the chirp is received by two antennas and multiplied together (mixed).

\item The resulting mixed signal ($x_{pq}(t)$) has constant frequency. Once again the frequency is proportional to the distance between antennas, and the angle of arrival.

\item We would like to take a spatial FFT of the mixed signals over all antennas at each sample time. Much like the Fourier transform in regular interferometry, this spatial FFT requires the signals to be sampled on a regular grid. To form a regular grid ($x'_l[n]$), we take the sampled, mixed signal from each pair of antennas ($x_{pq}[n]$) and average those products which have the same inter-antenna spacing ($l$), and therefore the same (and therefore redundant) geometric delays. This process is known as `gridding'. Gridding can also  be used to interpolate a non-uniform array geometry onto a uniform grid so that the FFT can be used.

\item The gridded signals from all antennas are comprise sort of `space-time tone'. I.e. for a given sample number ($n$), the spatial frequency of the tone is proportional to the angle of arrival ($\theta$). Similarly, for a given inter-antenna spacing ($l$), the temporal frequency is proportional to $\theta$.

\item For each sample number $n$, we take the DFT of the gridded signals over the spatial dimension (which can be implemented as an FFT). The result is DFT of a single tone ($X_k[n]$), which has a peak at the bin $k_0$.

\item Unlike with The Chirpolator, the peak in the DFT ($k_0$) is not a constant. In fact the peak increases linearly with the sample number $n$ and is proportional to the arrival direction $\theta$. As a result, a pulse of duration $T$ arriving from a direction $\theta$ will trace out a linear trajectory in time where it will cross a number of spatial DFT bins (Figure \ref{fig:Chimageator}) during its duration.

\item The angle of arrival and DM (equivalent to pulse duration $T$) are unknown. Therefore, at each sample time, we assume a set of trial angles and durations, which correspond to a set of trial trajectories. To form a detection metric $P(\theta)$ for each angle and duration, we sum along the trial trajectory (applying a phase correction $\Phi^*$ as we go).

\item Additional optimizations are possible. For example, the shorter trajectories can be calculated as partial sums along longer trajectories with the same gradient, and the spatial FFTs can be averaged before performing the trajectory sums. These optimizations are discussed in Section \ref{sec:chimg_optimizations}.

\end{enumerate}

\subsection{Formulation for a uniform linear array}
\label{sec:formulation_ula}

In this section, we develop a more rigourous description of The Chimageator.

Consider a linear, perfectly calibrated array of $M$ antennas, uniformly spaced with inter-element spacing $L$. If a linear chirp impinges on the array, the product of the signals from two antennas, indexed $p$, and $q$ is, therefore, given by:

\begin{eqnarray}
x_{pq}(t) & = & s_p(t)  s^*_q(t) \label{eq:xpq_chimg}\\
& = & s(t - p\tau) s^*(p - q \tau) \\
& =  & \exp{\left ( \pi j  \dot{f} (t - p\tau)^2 \right )} \exp{\left ( - \pi j  \dot{f} (t - q \tau)^2 \right )} \\
& = & \exp{\left ( - \pi j  \dot{f} \left (2 t \tau (p - q) \right ) + \tau^2(p^2 - q^2) \right )} \\
& \simeq & \exp{\left ( - 2 \pi j  \dot{f} t \tau \left  (p-q \right )  \right ) } \label{eq:chimg_xpq}
\end{eqnarray}

\noindent where

\begin{equation}
\tau = \frac{L}{c}\sin(\theta)
\end{equation}

As before, we can form the sampled signal by replacing $t \to n/f_s$. Next, we combine the values of $x_{pq}[n]$ for all baselines with the same spacing $l$, and the sample number $n$ in a process known as gridding. The uniform linear array has redundant spacings which can be combined and weighted according to:

\begin{eqnarray}
x'_{l}[n]  = \sum_{p=0}^{M - l - 1}{w_i x_{p, p + l} [n]}
\end{eqnarray}

\noindent where $l$ runs from 1 to $M-1$ (the auto-correlations are ignored). $ x'_{l}[n]$ corresponds to the visibility measured by combining all baselines with spacing $(l+1)D$ and $w_i$ are weights. $w_i=1$ corresponds to `uniform' weighting yielding the maximum signal-to-noise ratio but reduced resolution. $w_i = 1/(N-l)$ corresponds to natural weighting, yields maximal resolution but reduced signal-to-noise ratio. Through suitable choice of weights, an arbitrary array be interpolated onto a regular grid as required for the spatial FFT, by gridding with a spatially varying set of weights. The interested reader is referred to \citet[chapter 7, section 3]{Taylor99}, for details.

For a single chirp, we can substitute Equation \ref{eq:chimg_xpq} and assuming natural weighting and a uniform linear array, the gridded voltages simplify to:

\begin{eqnarray}
x'_{l}[n] & = &\sum_{p=0}^{M - l - 1}{w_i \exp{\left ( - 2 \pi j  \dot{f} n \tau \left  (p-( p + l) \right )  / f_s\right ) }} \\
& = &\exp{\left (  2 \pi j  \dot{f} n \tau l   / f_s \right) } \sum_{p=0}^{M - l - 1}{w_i } \\
& = &\exp{\left (  2 \pi j  \dot{f} n \tau l  / f_s \right) } \\
\end{eqnarray}

A spatial discrete Fourier transform of the gridded voltages yields:

\begin{eqnarray}
X_k [n] & = &  DFT\left \{ x'_l[n] \right \} \\
& = & \sum_{l=0}^{M - 1}{\exp{ \left ( \frac{-2 \pi j k l}{N} \right )} \exp{\left (  2 \pi j  \dot{f} \tau l \frac{n}{f_s} \right ) }  } \\
& = & \sum_{l=0}^{M-1}{2 \pi j l \left ( \frac{ -k }{N} + \tau \dot{f} \frac{n}{f_s}  \right )} \\
& = & D_M(k - k_0) \Phi(k - k_0)
\end{eqnarray}

\noindent where $D_M(f)$ is the Dirichlet kernel defined earlier, and $\Phi$ is a unit-amplitude phase term. This spatial DFT can be efficiently implemented as a Fast Fourier transform.

\begin{figure}
\centering
\includegraphics[width=\linewidth]{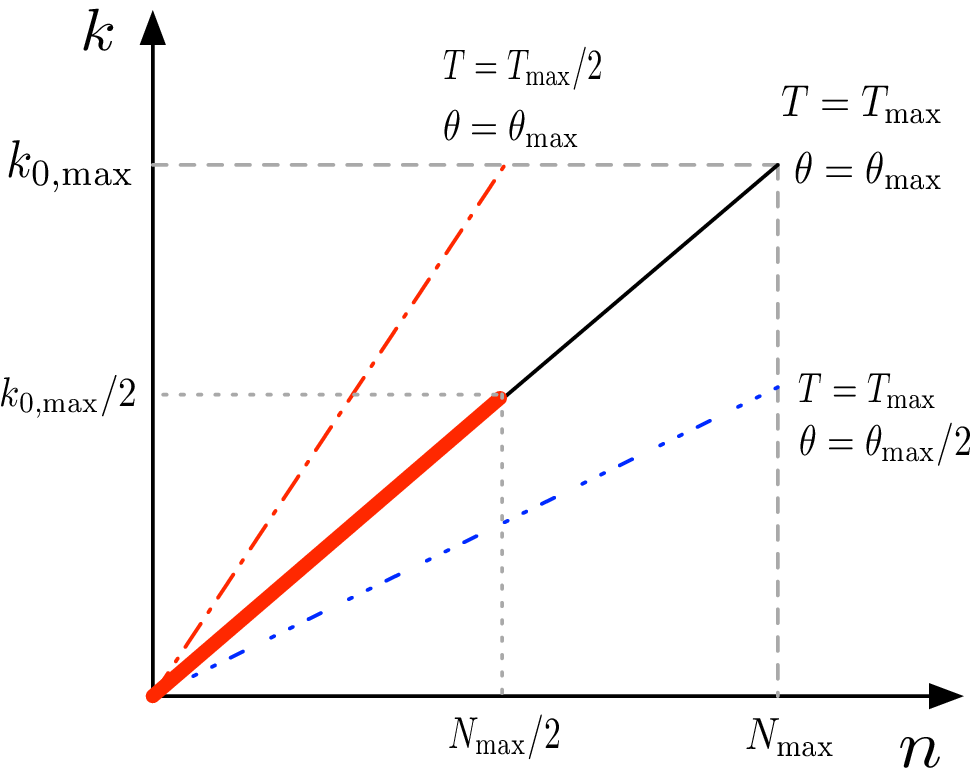}
\caption{Trajectories of linear chirps with varying durations ($T$) and angles of arrival ($\theta$) after gridding and Fourier transforming with The Chimageator. As the sample number ($n$) increases, the peak of the DFT ($k_0$) increases linearly (see Equation \ref{eq:chimgf0}). Two types of trajectory are shown with dashed lines: The $\theta_{\rm max}$ case, which corresponds a range of dispersion delays, and a single arrival angle at the edge of the field of view; and the $T_{\rm max}$ case, where each trajectory corresponds to the longest dispersion of interest and a range of arrival angles. The thick line is the trajectory corresponding to $T_{\rm max}/2$ and $\theta_{\rm max}/2$, which lies along the $T_{\rm max}, \theta_{\rm max}$ trajectory and can be therefore be computed from the partial sums along the $T_{\rm max}, \theta_{\rm max}$ trajectory.}
\label{fig:Chimageator}
\end{figure}

A chirp crossing a bandwidth $B=f_s$ in time $T$, arriving from angle $\theta$  signal puts power in the DFT bin given by:

\begin{eqnarray}
k_0 (n, T, \theta) & = & \dot{f} \tau M \frac{n}{f_s} \\
& = & \frac{n}{T}  \tau M\\
& = &\frac{n}{T} \frac{ML}{c} sin(\theta) \\
& \simeq &\frac{n}{T} \frac{b_{\rm max}}{c} \sin(\theta) \label{eq:chimgf0}
\end{eqnarray}

\noindent which is very similar to the expression for The Chirpolator described previously, with the key difference that in this case, the $k_0$ term now depends linearly with sample number $n$ rather than baseline length $b_{pq}$. Thus, a chirp signal will appear as power along a diagonal trajectory in $n$-$k$ space, as shown in Figure \ref{fig:Chimageator}.

For a chirp beginning at $n = 0$, the trajectory ends at the DFT bin given by:

\begin{eqnarray}
k_{\rm 0, end} & = &k_0 ( T f_s, T, \theta) \\
& = & f_s \frac{b_{\rm max}}{c} \sin(\theta)
\end{eqnarray}

To form an image, we can sum across the diagonal trajectory in DFT bins and time, applying the inverse of the phase term to produce an intensity image for a given dispersion delay, by:

\begin{eqnarray}
P_T(\theta) & = & \sum_{n=0}^{f_s T }{\Phi^*\left (k_0(n, T, \theta) \right ) X_k \left (k_0(n, T, \theta) \right )} \label{eq:chimg_pth}
\end{eqnarray}

\noindent with the scalar energy computed as in Equation \ref{eq:eth}.

\section{Method of comparison}

We have described two new antenna-coherent techniques for detecting dispersed pulses with interferometers. In this section, we describe our method of comparing our techniques to two existing classical techniques with roughly equivalent sensitivity: Fourier imaging, and direct beamforming with frequency incoherent processing (see Figure \ref{fig:taxonomy}). These classical techniques are described in Sections \ref{sec:direct_beamforming},  \ref{sec:fourier_imaging} and \ref{sec:incoherent_dedispersion} respectively.

Ideally, we would like to compare the techniques in terms of the true implementation costs. But, evaluating the true implementation cost is complicated by a number of considerations:

\begin{itemize}

\item The choice of survey parameters (e.g. minimum \& maximum DM, center frequency).

\item The telescope parameters (e.g. number of antennas, system bandwidth, baseline distribution).

\item The economics of available technologies.

\item The details of the implementation on a given technology. For example, how an algorithm is parallelized over a number of processors.

\item The techniques do not yield equivalent sensitivities in certain situations (e.g. Section \ref{sec:ext_millisec}).

\item The parametrization of the algorithms themselves.

\end{itemize}

To explain the final point further: an implementation of an algorithm requires a set of parameters that affects both the cost and the sensitivity of that implementation (e.g. number of channels for interferometric imaging, or $F$ for The Chirpolator). For each algorithm, the relationship between the parameters and sensitivity is complicated, and there is no straightforward way to choose realizations that yield equivalent sensitivities for all techniques so that their costs can be compared fairly.

\subsection{A simple model for evaluating algorithm cost}

To help illustrate, in very approximate terms,  the differences in operations and data rates required by different methods, we propose a simple model.  In this model, we split each algorithm up into two basic functional blocks: the processing required before an integrate-and-dump step, and the processing required after it. We also consider the data rate required between the two blocks, i.e. immediately after the integrate-and-dump step. We acknowledge that this model does not consider very important details of how data is transported within each block, and acknowledge that the bandwidth bottlenecks may indeed be within each block, rather than between the two. But, the bandwidth requirements inside each block are a strong function of the way the the processing is parallelised inside each block, and quantifying the many different methods for doing this parallelization are outside the scope of this paper.

This functional breakdown applies to the techniques as follows:

\begin{description}

\item[The Chirpolator]: The pre-integrator step is the sliding-DFT (Section \ref{sec:sliding}). The integrated output is a set of DFT results per DM trial. The post integrator step is the imaging per DM trial. Detail of the data and operations rates are described in Appendix \ref{sec:app_chirpolator}.

\item[The Chimageator]: The pre-integrator steps include gridding and integration to the shortest sampling interval. The integrated output is a sequence of partially-averaged images. The post-integrator steps include the remaining integration for the full range of DM trials, and the imaging. Detail of the data and operations rates are described in Appendix \ref{sec:app_chimg}.

\item[Fourier Imaging]: The pre-integrator steps include cross-correlation and integration. The integrated output is the visibilities. The post-integrator steps  includes gridding, FFT and tree incoherent dedispersion \citep{taylor1974ddisp}.  For the the bandwidth requirement, we sum both the requirements for both the visibilities, and the `corner turn' required for dedispersion. Operations rates are described by \citet{Cordes97}

\item[Direct Beamforming]: We form as many tied array beams as required to cover the entire the primary beam. The pre-integrator steps include the beam forming and integration. The integrated output is a power spectrum per beam. The post integrator step is tree incoherent dedispersion. Operations rates are described by \citet{Cordes97}.

\end{description}

\subsection{Array, survey and algorithm parameters}

To arrive at concrete values of bandwidth and operations rate, we must define a full set of parameters for an array, survey and each algorithm. To motivate our example, we choose a set of parameters based on the SKA from \citet{Cordes97}, as shown in Table \ref{tab:survey_params}. Clearly, the evaluating the performance of all techniques as a function of all parameters results in a highly-multidimensional dataset. For the sake of simplicity, we leave only one free parameter: the number of antennas in the array ($M$). We let $M$ go from 2 antennas up to 2000 antennas, which covers the range of values for SKA and its pathfinders.

\begin{table}
\caption{Parameters used in our example model. The parameters for The Chirpolator and The Chimageator are defined in Appendices \ref{sec:app_chirpolator}, and \ref{sec:app_chimg} respectively.
}
\label{tab:survey_params}
\centering
\footnotesize
\begin{tabular}{l r }
\hline\hline
Parameter & Value \\
\hline
Array  parameters \\
System Bandwidth  (MHz) & 400 \\
Antenna size (m) & 12 \\
Maximum Baseline (m) & 1000 \\
Center Frequency (GHz) & 1.4 \\
Number of polarizations & 2 \\

\hline
Survey parameters \\
Minimum DM (cm$^{-3}$pc) & 10 \\
Maximum DM (cm$^{-3}$pc) & 1000 \\

\hline
Fourier imaging \& direct beamforming \\
\hline
Number of frequency channels & 1000 \\
Number of DM trials & 1000 \\
Integration time (seconds ) & $10^{-4}$ \\
Bytes per visibility (post correlator) & 2 \\
Bytes per image pixel & 1 \\
\hline
Chirpolator specific \\
\hline
DM step ($\epsilon$)  & 0.1 \\
Smearing support size (F) & 1 \\
Time oversampling ($\kappa_t$) & 4 \\
Spatial oversampling ($\kappa_s$) & 1 \\
Bytes per DFT bin & 2 \\
\hline
Chimageator specific \\
\hline
DM step ($\epsilon$) & 0.1 \\
Smearing support size (F) & 1 \\
Time oversampling ($\kappa_{t, 0}$) & 4 \\
Spatial oversampling ($\kappa_s$) & 1 \\
Operations per grid point & 50 \\
Bytes per FFT bin & 2 \\
\hline
\end{tabular}

\end{table}

\section{Results}

\begin{figure}
\centering
\includegraphics[width=8cm]{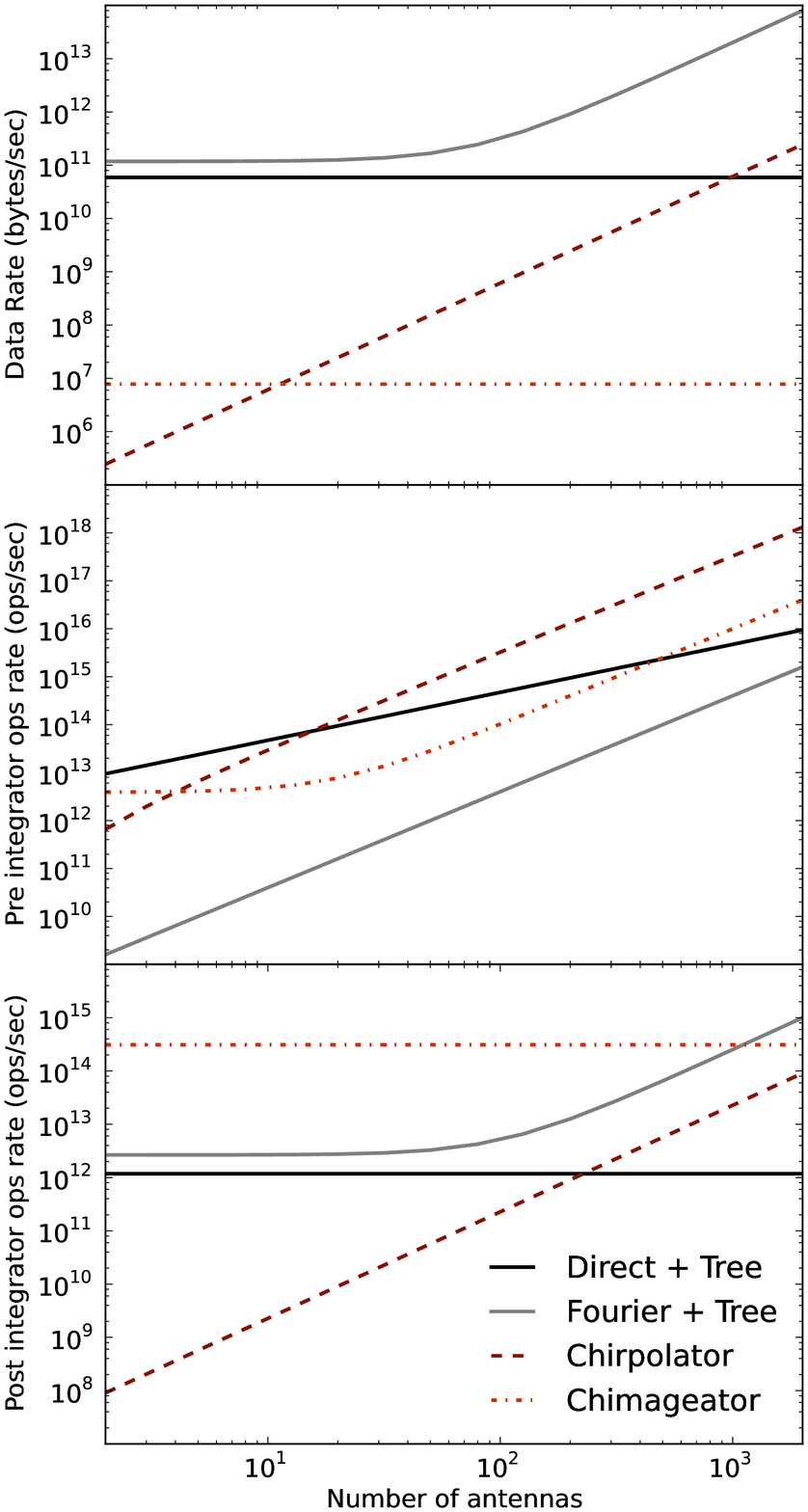}
\caption{Data and operations rates for a dispersed pulse survey as a function of algorithm and number of antennas. Survey parameters are given in Table \ref{tab:survey_params}. Top Panel: data rate between pre-integrator and post-integrator steps. Middle panel: operations rate before the integrator. Middle panel: operations rate after the integrator.  `Fourier + Tree' and `Direct + Tree' signify Fourier imaging and Direct beamforming respectively to form beams, and using tree dedispersion described by \citet{taylor1974ddisp} for dedispersion.}
\label{fig:askap_compcost}
\end{figure}

For all but the largest arrays, The Chimageator and The Chirpolator have substantially superior bandwidth requirements than the classical techniques (Figure \ref{fig:askap_compcost}). The lower bandwidth requirements are achieved because of a difference in timescale that needs to be sampled by the integrate-and-dump step:  our techniques sample the shortest dispersion delay, while the classical techniques sample the shortest dedispersed pulse duration. As a dedispersed pulse can be substantially shorter than the dispersion delay, the classical techniques must dump their integrators at a much higher rate, therefore requiring larger bandwidth between the functional blocks. One additional factor worsens the bandwidth requirements for Fourier imaging in particular: below about $\simeq 100$ antennas the bandwidth dominated by the dedispersion  `corner turn'.

In terms of post-integrator operations rate, The Chirpolator betters all other techniques up to $\simeq 200$ antennas. This low rate for small arrays is consequence of both the low input bandwidth, and the fact the imaging operates on a per-baseline basis. Above $\simeq 200$ antennas, the Direct beamforming method is the clear winner, as the dedispersion cost is fixed by the longest baseline, rather than the number of antennas.

In terms of pre-integrator operations rate, Fourier imaging is clearly the most efficient for all array sizes of interest, with our techniques requiring between 2 and 4 orders of magnitude more operations for equivalent array sizes.

\section{Discussion}

For any array size, there is no clearly superior algorithm in all measures. The Chirpolator has high pre-integrator operations rate, but has good post-integrator and data requirements for small to medium arrays. The Chimageator has consistently high a post-integrator operations rate. Fourier imaging is computationally attractive but has very high data rate requirements, either due to the corner turn in small arrays, or the visibility data rate in large arrays. Direct beamforming has very high operations and data rate requirements for small arrays but becomes somewhat competitive for larger ones. The preferred algorithm, therefore, will depend on the details of the array, survey and algorithm parameters, and the economics of available computing technologies.

The economics of computing technology are changing rapidly. The increase in arithmetic capability of processors has been well described by Moore's law; that is, the number of transistors (and by inference, arithmetic capability) on a chip doubles every 18 months. While this prodigious improvement is very welcome for the arithmetic part of the problem,  it does not hold for data rate, which has traditionally grown much more slowly. We propose that, because the arithmetic capability of processors is outstripping the bandwidth capability, our techniques with their superior data rate performance, will become more and more favorable as technology progresses in spite of their requirements for higher operations rate. Therefore, in the time scale of the SKA, its pathfinders, our techniques may be preferred over the classical ones.

\subsection{Further work}

\subsubsection{Effect of calibration errors}
\label{sec:calibration}

In our analysis we have assumed an ideal, perfectly calibrated array, in which all the antenna gains are equal and have zero relative phase. In practice, each antenna will have uncalibrated errors in gain and phase which will affect the performance of our algorithms. While a detailed discussion of the effect of calibration errors is outside the scope of this paper, we present here a simple proof that phase errors (which we model as delay errors) in The Chirpolator case, will result in decoherence across the array and reduced SNR.

If we assume the uncalibrated delay error between two antennas is $\tau_{\rm err}$ then we can make the substitution $\tau \to \tau + \tau_{\rm err}$ into Equation \ref{eq:f0Bt} to obtain the frequency of the tone after mixing:

\begin{eqnarray}
\hat{k}_0 & = & B(\tau + \tau_{\rm err}) \\
& = & k_0 + k_{\rm err}
\end{eqnarray}

Therefore, a delay error changes the frequency of the mixed signal, and shifts the entire DFT spectrum from $X_{pq}[k]$ to $X_{pq}[k + k_{\rm err}]$. The shift in the DFTs reduces the amplitude of the detection metric, which is formed by a vector sum of the phase-corrected DFT bins from each antenna pair. The detection metric has a maximum value when all the phase-corrected DFT bins have the same absolute phase. If an antenna pair contains a delay error, the each phase-corrected DFT bin will not have the same absolute phase, and the vector sum will not be over a straight line (Figure \ref{fig:calerrors}), resulting in reduced amplitude of the sum. This process  can be quantified by substituting $\hat{k_0}$ into Equation \ref{eq:pth}:

\begin{figure}
\centering
\includegraphics[width=\linewidth]{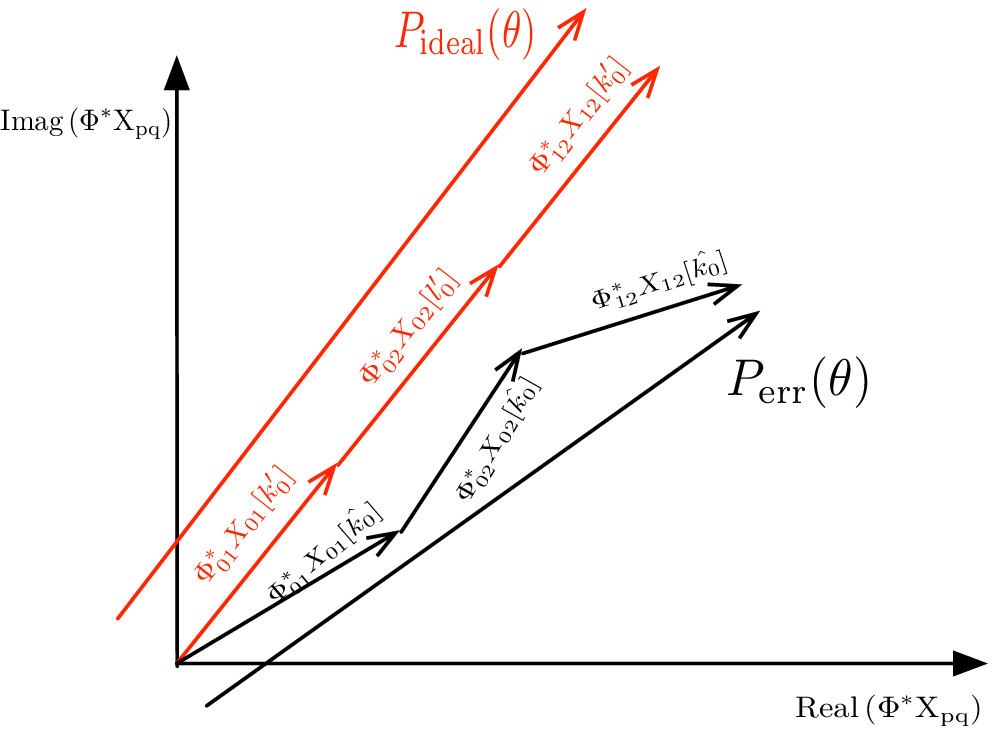}
\caption{Errors in delay calibration reduce the amplitude of the detection metric. Here we plot the formation of the detection metric $P(\theta)$ as the vector (i.e. complex) sum of the phase corrected DFT results from three antenna pairs. In the ideal case (red), the phase correction ($\Phi^*_{pq}$) perfectly corrects for the known phase in the DFT bins ($X_{pq}$), and each result has the same absolute phase. The resulting detection metric ($P_{\rm ideal}(\theta)$), is fully coherent. If delay errors are present, each DFT bin has a residual phase that is different for each antenna pair. The resulting detection metric ($P_{\rm err}(\theta)$) has a smaller amplitude, because the vectors do not add into a straight line.}
\label{fig:calerrors}
\end{figure}

\begin{eqnarray}
P_{\rm err}(\theta)  & = & \sum_{p=0}^{M-1} \sum_{q=p+1}^{M-1} \Phi^*_{pq}(k_0' - k)  X_{pq}[\hat{k_0}] \\
& = & \sum_{p=0}^{M-1} \sum_{q=p+1}^{M-1} \Phi^*_{pq}(k_0' - k)  \Phi_{pq}(k_0' +  k_{\rm err} - k_0) D_N(k_0' + k_{\rm err} - k_0) \\
& = & \sum_{p=0}^{M-1} \sum_{q=p+1}^{M-1} \exp \left (j \pi (k_0' - k) \right) \exp \left (-j \pi (k_0' +  k_{\rm err} - k_0 \right ) D_N(B \tau_{\rm err} ) \\
& = & \sum_{p=0}^{M-1} \sum_{q=p+1}^{M-1} \exp \left (-j \pi B \tau_{\rm err} \right )D_N(B \tau_{\rm err} ) \\
& \leq & P_{\rm ideal}(\theta)\label{eq:pth_calerr}
\end{eqnarray}

The inequality in Equation \ref{eq:pth_calerr} is a result of the triangle inequality for vector addition (See Figure  \ref{fig:calerrors}), and the fact that $D_N(x) \leq D_N(0)$.

It is clear from this argument that delay errors will result in a reduced detection metric, resulting in a drop in signal-to-noise ratio. We leave a quantitative analysis of this effect, and other calibration effects for future work.

\subsubsection{Extension to millisecond pulsars}
\label{sec:ext_millisec}
Both our methods have assumed that a chirp is received in isolation, meaning that during the duration $T$ of a chirp, no other chirps are received. This condition is violated for millisecond pulsars, which have short periods and can have large DMs. The combination of short period and large DM means that a chirp will not have finished traversing the system bandwidth $B$ before a subsequent chirp is received.

We can write the isolated chirp condition for a pulsar with period $P$ as:

\begin{eqnarray}
P & > & T \\
& > & \mu DM (\nu_1^{-2} - \nu_2 ^{-2}).
\end{eqnarray}

If the isolated chirp condition is not satisfied, there are multiple chirps occupying the bandwidth at any one time. These additional chirps produce additional mixing products at the multiplication steps (i.e. in Equation \ref{eq:xpq} and \ref{eq:xpq_chimg}) which appear at frequencies that are outside the frequencies searched in the isolated chirp case. If only isolated chirp processing is performed, the energy in the additional mixing products is effectively lost, with a resulting loss in SNR. Our techniques will still operate effectively, but the SNR achieved will not be as high as with processed by other methods. Quantifying loss of energy to mixing products, and resulting loss in SNR, is outside the scope of this paper.

\begin{figure}
\centering
\includegraphics[width=\linewidth]{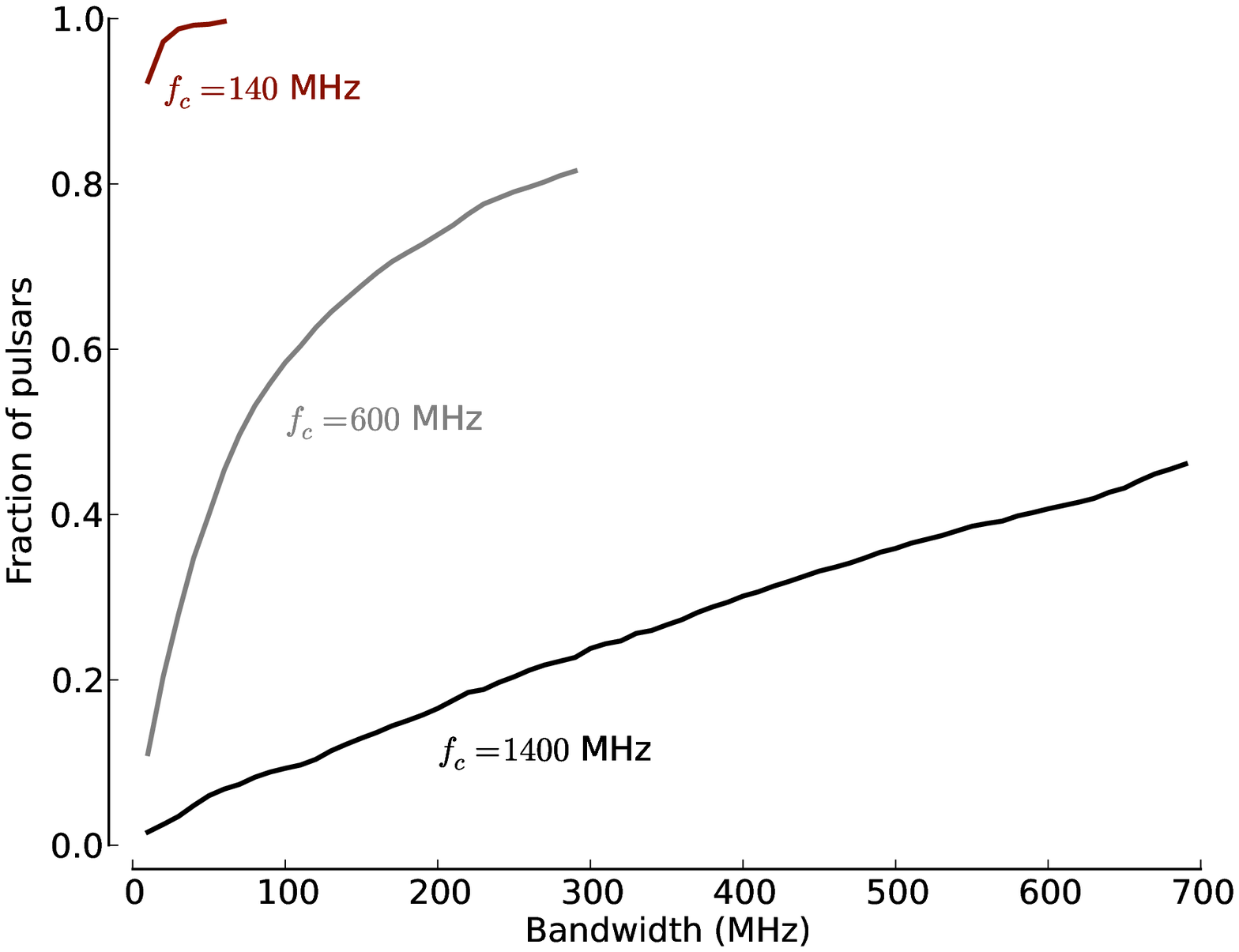}
\caption{Fraction of known pulsars that violate the isolated chirp condition, as a function of system bandwidth and center frequency ($f_c$). Known pulsars are taken from the ATNF pulsar catalog.}
\label{fig:frac_violators}
\end{figure}

To determine what fraction of pulsars violate this condition, we use the ATNF pulsar catalog \citep{Manchester05}\footnote{http://www.atnf.csiro.au/research/pulsar/psrcat}. This catalog contains the DM and period for all known pulsars. At a bandwidth of 400~MHz at 1.4~GHz, 30\% of the known pulsars have periods that are too high to satisfy the isolated chirp condition for their DM (Fig. \ref{fig:frac_violators} ), indicating this effect is important.

\section{Summary}

We have described two new techniques for detecting dispersed pulses with radio interferometers, which we call The Chirpolator and The Chimageator. These techniques have antenna-coherent sensitivities in the isolated chirp case and substantially lower data rate requirements than other coherent methods for realistic array configurations including the SKA and its pathfinders. For small to medium array sizes The Chirpolator is also more efficient than classical techniques in terms of post-integrator operations rate. While the pre-integrator operations rates our methods high in some cases, the changing economies of computer design may flavor lower bandwidth requirements of our new techniques in spite of their high operations rate requirements.

\begin{acknowledgements}
KB acknowledges the support of an Australian Postgraduate Award and a CSIRO top-up scholarship. The Centre for All-sky Astrophysics is an Australian Research Council Centre of Excellence, funded by grant CE11E0090. We thank the anonymous referee for their prompt and helpful comments.
\end{acknowledgements}

\bibliographystyle{apj}
\bibliography{Master.bib}

\appendix

\section{The Chirpolator: Analysis and implementation}
\label{sec:app_chirpolator}
In this appendix we describe additional extensions to The Chirpolator to include multiple telescopes in 3D and non-linear dispersion. We describe novel methods for efficiently implementing The Chirpolator and also derive equations for the resolution and data and operations rate requirements.

\subsection{Multiple telescopes in 3D}

The generalization  to arbitrary arrays of elements in three dimensions is most easily done in the notation of interferometry \citep[Chapter 2]{Taylor99}.

If we measure the $[u, v, w]^T$ baseline vector in units of distance (not wavelength), then the geometric delay for a 3 dimensional array is:
\begin{equation}
\tau = \frac{ul + vm + w \left (\sqrt{1 - l^2 - m^2} - 1 \right )}{c} \label{eq:tau3D}
\end{equation}

\noindent where $l$ and $m$ are the direction cosines in the $u$ and $v$ directions respectively. $l$ and $m$ define the angle of interest analogous to $\theta$ in the 1D case.

The method of computing the intensity image then proceeds in much the same manner, with $k_0$ computed with equations \ref{eq:f0Bt} and \ref{eq:tau3D}, and with $P(\theta)$ evaluated over two angular dimensions instead of one.

\subsection{Non-linear dispersion}
\label{sec:nonlinear_dispersion}

In the main text, beginning at Equation \ref{eq:chirp}, we have assumed a linear chirp. In fact, at most frequencies and bandwidths of interest (i.e. below 10~GHz, and bandwidths $>$100~MHz), the cold plasma dispersion law is much more accurately modeled as $\propto \nu^{-2}$ as shown in Equation \ref{eq:dmdelay} and Figure \ref{fig:delayvsfreq}. In this section we describe the effect of the true dispersion law on Chirpolator processing (decoherence), and propose a solution (oversampling).

\subsubsection{The Problem: Decoherence in the DFT bins}

\begin{figure}
\centering
\includegraphics[width=\linewidth]{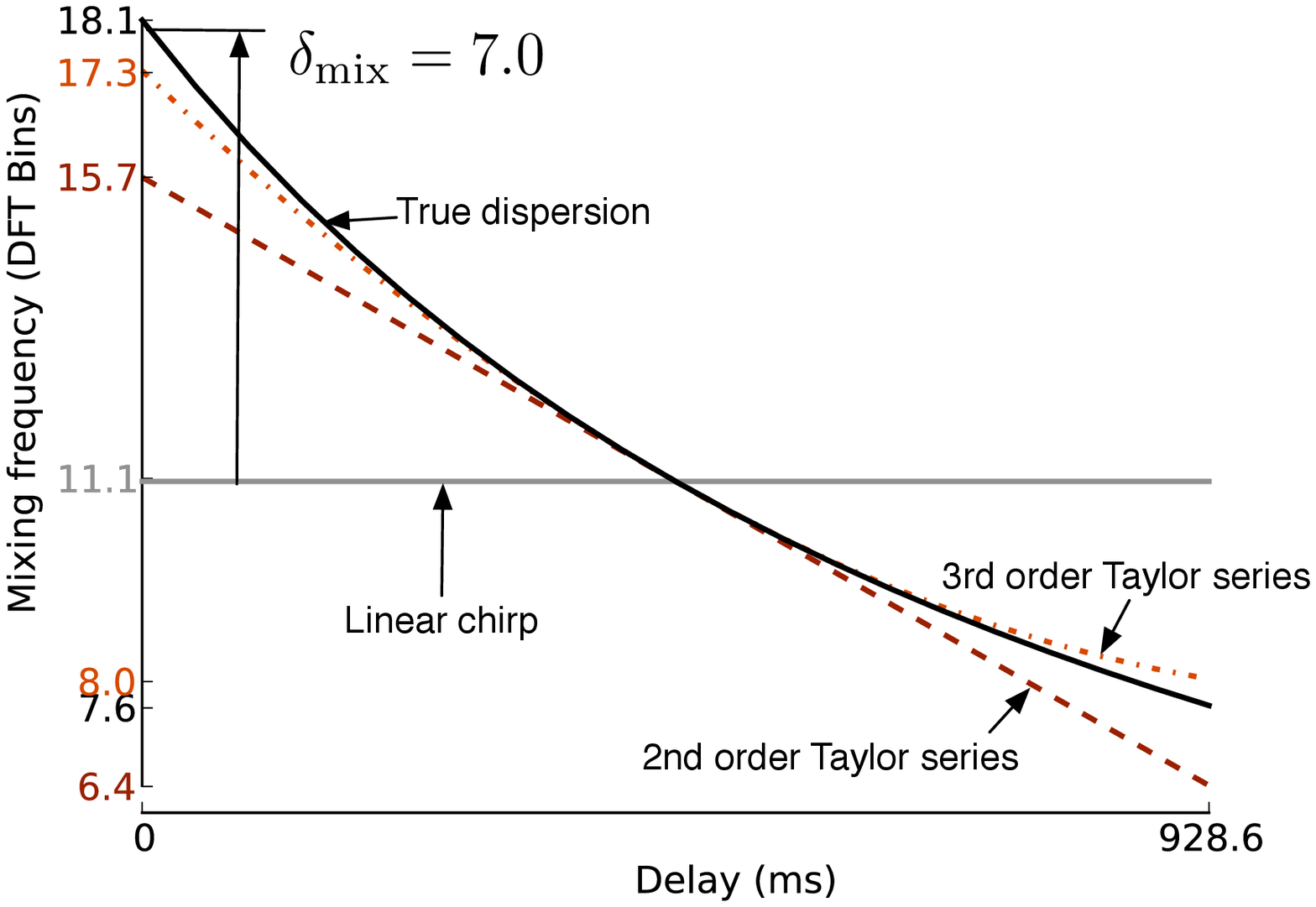}
\caption{The simulated mixing frequency  as a function of time for a single antenna pair of The Chirpolator (see Equation {eq:approxerror1}). A range of approximations are shown. The parameters for this simulation were: a DM of $100 \unit{cm^{-3} pc}$ and a bandwidth of 400~MHz centered at 1.4~GHz, $\theta=0.5$ degree and a baseline of 1~km.}
\label{fig:approx_error}
\end{figure}

\begin{figure}
\centering
\includegraphics[width= \linewidth]{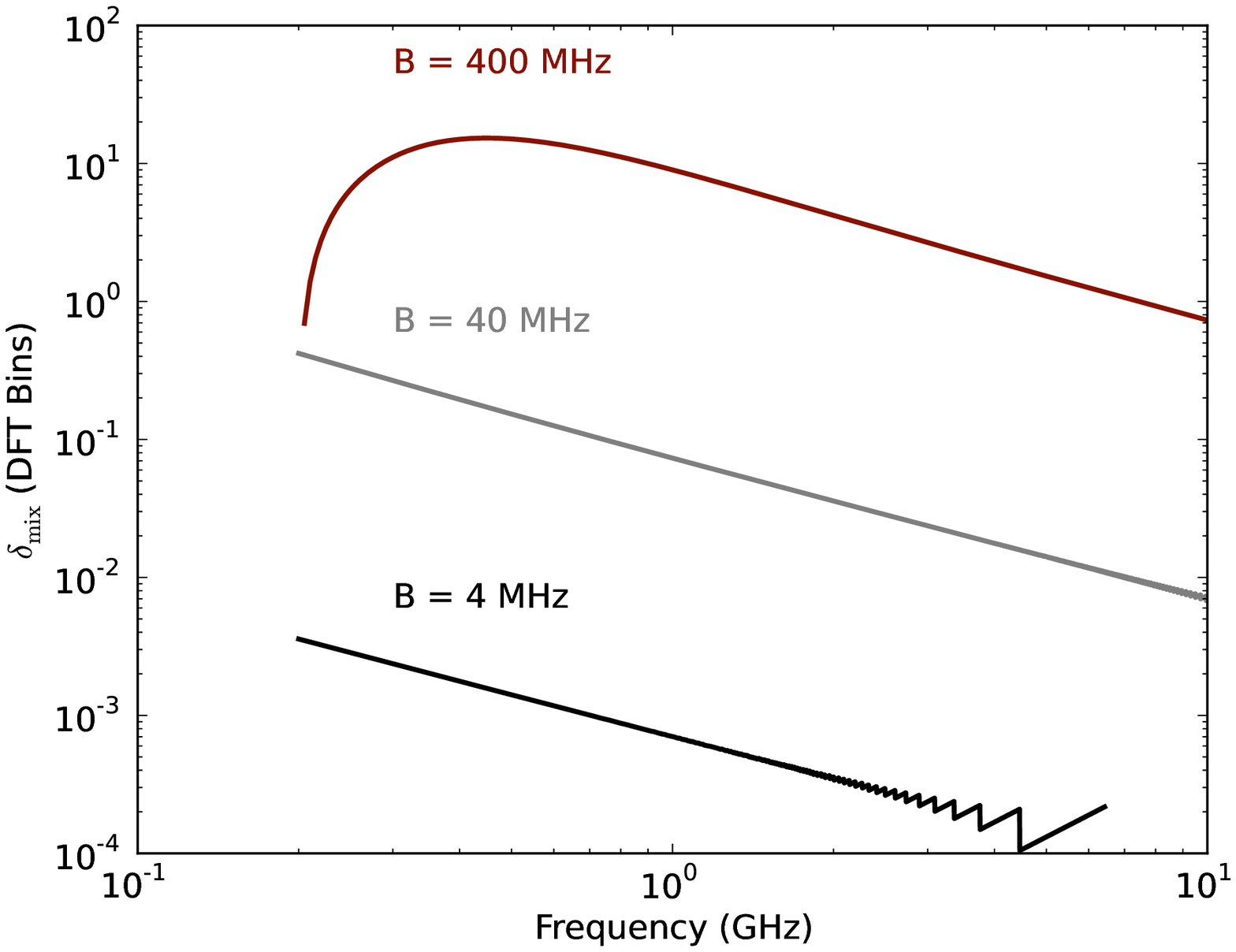}
\caption{Error in mixing frequency $\delta_{\rm mix}$ as a function of center frequency and bandwidth ($B$), assuming DM of $100 \unit{cm^{-3} pc}$, 1~km baseline and $\theta=0.5$ degree.}
\label{fig:nonlinerr_vs_freq}
\end{figure}

To determine the effect of the higher order terms on Chirpolator processing, we begin by considering the frequency of the mixed signal $x_{pq}$ (Equation \ref{eq:xpq}), which is the difference between the instantaneous frequencies of the signals from the two antennas. Assuming a delay $\tau \ll T$, the instantaneous frequency difference between the two chirps is given by:

\begin{eqnarray}
\nu_{\rm mix}(t) & = & \nu_2(t - \tau) - \nu_2(t) \\
& \simeq & -a_1 \tau - 2 a_2 \tau (t - T/2) \label{eq:freqdiff}
\end{eqnarray}

\noindent where we have used the Taylor expansion described in Equation \ref{eq:freqvst2}. From Equation \ref{eq:freqdiff} we can see that the effect of nonlinear dispersion on The Chirpolator processing is to smear out the signal across a wider range of frequencies after mixing the two antenna signals (Fig. \ref{fig:approx_error}). The departure of the frequency from the linear assumption is significant for typical array configurations and dispersion (Fig. \ref{fig:nonlinerr_vs_freq}), and is worst far from the phase centre, on the long baselines, and at $t = 0$, where it can be approximated as the difference between the linear approximation and the 3rd order Taylor series (Equation \ref{eq:freqdiff}):

\begin{eqnarray}
\delta_{\rm mix} & = &\tau T^2 \left (a_2  - \frac{3}{4} a_3 \tau T \right )\label{eq:approxerror1}
\end{eqnarray}

When a signal with non-constant frequency is passed through a DFT, the amplitude of the DFT output is reduced, which we call decoherence.

We identify three regimes in which the system operates:

\begin{itemize}

\item The smearing is  $\ll 1$ bin, in which case the decoherence is small and can be ignored.

\item The smearing is $\sim 1$ bin, in which case the signal still occupies only one bin, but the decoherence within that bin is significant. In this case, the DFT must be broken into a number of sub-integrations, with each sub-integration requiring a complex phase rotation to recover the coherence.

\item The smearing is $> 1$ bin, in which case there is energy in multiple bins. The DFTs must be broken into a number of sub-integrations. The final output must before formed with is a complex phase rotation of a range of sub-integrations of \emph{different} DFT bins.
\end{itemize}

If the smearing is $>1$ bin, (e.g. Fig.  \ref{fig:approx_error}), the true dispersion occupies a higher DFT bin than the linear assumption for approximately half the pulse duration. To capture energy from the higher frequencies, additional DFTs must be computed that would not be required under the linear assumption. In the nonlinear case, the maximum number of DFT bins increases from $k_{0, max}$ to $k_{0, max} + \delta_{\rm mix}$, which increases the operations and data rate requirements for the DFT step. At the worst case longest baseline of 1~km, at 1.4~GHz, 400~MHz and  0.5 degrees from the phase centre, $\delta_{\rm mix} = 7$, and the number of DFT bins required increases by a factor of $ \delta_{\rm mix}/k_{\rm 0, max} = 63\%$. The additional DFT bins increases non-linearly as a function of baseline, so accurately estimating the total increase over the whole array requires a knowledge of the exact baseline distribution. To obtain an approximate figure, assuming a baseline distribution where the mean baseline length is half the maximum baseline length, we propose that the total increase is approximately half the worst case figure, i.e. 32\%.

\subsubsection{The solution: oversampling}
The key to handling the nonlinear dispersion, therefore, is to dump the integrator more often than required for the nonlinear case (oversample), and phase-correct the results to obtain the coherence again. To quantify the amount of oversampling required where the smearing is $>1$ bin we need to quantify the response of the DFT to the mixed, non-linearly dispersed signal. As shown in Fig \ref{fig:approx_error}, the frequency of the mixed signal is well approximated by the third order Taylor expansion described in Equation \ref{eq:freqvst2}. The phase of the mixed signal is, therefore, given by the integral:

\begin{eqnarray}
\phi_{\rm mix}(t) & = & \int_{0}^{t}{2 \pi \nu_{\rm mix}(t') dt'} \\
& = & 2 \pi (-a_1  \tau  t - a_2  \tau  t (t - T)) \\
& = & 2 \pi ( t (-a_1 \tau + a_2 \tau T) - t^2 a_2 \tau).
\end{eqnarray}

The sampled, mixed signal can then be expressed as:

\begin{eqnarray}
m[n] & = & \exp(j \phi_{\rm mix}[n]) \\
& = & \exp \left [ 2 \pi j \left (   \frac{n}{f_s} (-a_1 \tau + a_2 \tau T) - \left( \frac{n}{f_s} \right)^2 a_2 \tau \right) \right ]
\end{eqnarray}

and we take the DFT over $N = f_s T$ samples to obtain:

\begin{eqnarray}
X_m[k] & = & \sum_{n = 0}^{N-1}{\exp \left(-2 \pi j k n /N \right) m[n]  }\\
& = & \sum_{n = 0}^{N-1}{\exp 2 \pi j \left[ n(-a_1 \tau + a_2 \tau T - k/N) + n^2 (-a_2 \tau /f_s^2)\right] }\label{eq:mf}
\end{eqnarray}

The term that is linear with $n$ has already been dealt with in Equation \ref{eq:phidn}, and is simply the DFT of a single tone, so we turn our attention to the $n^2$ term and define the sum:

\begin{eqnarray}
G(a, L) = \sum_{n = 0}^{L - 1}\exp\left(2 \pi j a n^2\right) \label{eq:gaussum}
\end{eqnarray}

\noindent where:

\begin{eqnarray}
a = -a_2 \tau /f_s^2
\end{eqnarray}

From Equation \ref{eq:gaussum} it is clear that $G(0,L) = L$, and that for non-zero values of $a$, the $a n^2$ term introduces oscillations, effectively moving the instantaneous frequency into the adjacent DFT bins, so that $|G(a, L)| < L$ for non-zero $a$.

We want to determine how large L can be made before some fraction of the energy will be lost to adjacent DFT bins. Equation \ref{eq:gaussum} defines the result of summing a chirp with an initial instantaneous frequency of zero, which is essentially the centre of the DFT bin. If we define the the coherence loss, or loss in amplitude as:
\begin{eqnarray}
\eta = |G(a,L)|/L
\end{eqnarray}

Therefore, the value of $L$ that maintains a required $\eta$ is the number of samples to traverse half the DFT bin and maintain a given loss. To calculate the required oversampling for chirp that crosses and entire DFT  bin, we can pose the question: what oversampling factor $\kappa_t = N/2L$ is required to maintain $\eta(a, L)$ above a specified threshold?

\begin{figure}
\centering
\includegraphics[width= \linewidth]{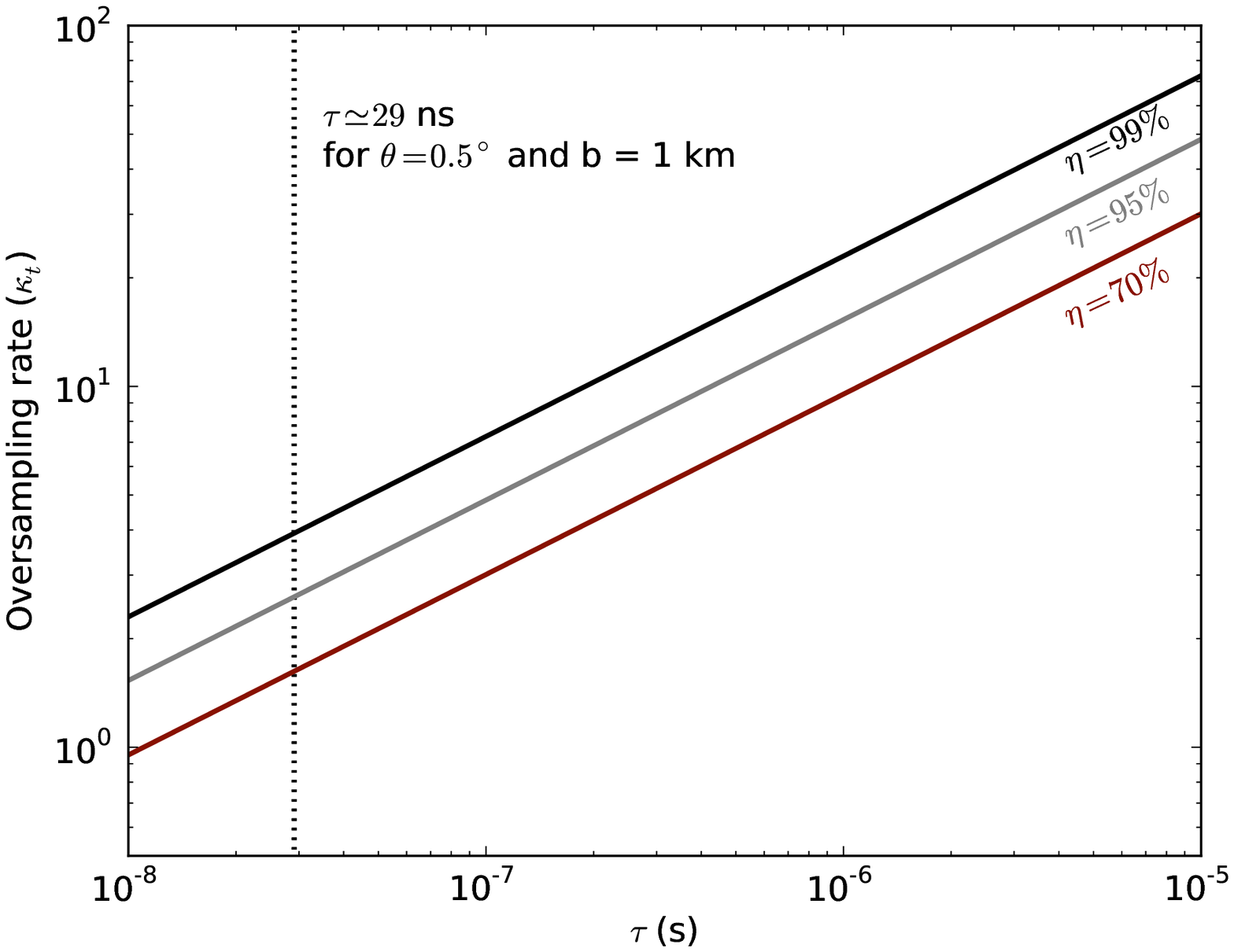}
\caption{One method for handling nonlinear dispersion with The Chirpolator is to increase the oversampling rate. Above is the required oversampling rate ($\kappa_t$) for a nonlinear chirp vs the geometric delay ($\tau$) for single a baseline operating at $f_c=1.4$~GHz, B=400~MHz and a range of different coherence losses ($\eta$). The vertical dashed line is the geometric delayfor $\theta=0.5^{\circ}$ and a baseline of 1~km. The required oversampling rate is independent of DM.}
\label{fig:os_vs_tau_loglog}
\end{figure}

Do get an approximation of the required oversampling factor, we have simulated a typical case for the SKA, with a case with a one-sided frequency smearing of the order of 7 DFT bins, which reasonably large in the context of Figure \ref{fig:nonlinerr_vs_freq}. We conclude that a 4 times oversampling yields $\eta \simeq 99\%$ (Fig.  \ref{fig:os_vs_tau_loglog}). We have found empirically that the required oversampling is independent of DM.

Because the time of arrival is not known, a one would typically require $\simeq 4$ times oversampling to obtain a sample which is integrated over a large fraction of the incoming signal. The equivalence of the oversampling rates required for time oversampling, and nonlinear dispersion correction, implies that nonlinear dispersion does not substantially drive the oversampling in this instance.

\subsection{Implementation optimizations}
\label{sec:implementation_optimizations}

In the main text we assumed a single value of $\dot{f}$ (equivalently a single value of the DM), and that the DFT window is exactly time-aligned with the chirp. In practice, neither the time of arrival for the chirp, nor the $\dot{f}$ are known in advance and we would like to maximize our chances of finding the signal. The maximum likelihood approach to the problem of maximizing the detection probability when the waveform parameters are unknown, is to pass the signal through many different matched filters, each with a particular realization of the unknown parameters. In our case, we would evaluate $P(\theta)$ and $X_{pq}[k]$ independently on a range of values of $\dot{f}$ and on a set of overlapping windows in time.

Significant computational savings can be made as described in the following sections.

\subsubsection{Compute only DFTs required on a baseline basis}
\label{sec:only_dft_baselines}

We do not have to compute the same number of DFT bins for each pair of antennas. In fact, for a given pair of antennas, we only have to compute the DFT for values of $k$ up to approximately $k_{\rm 0,max}$, as illustrated in Fig. \ref{fig:freqvsbaseline}. With values from typical radio telescopes,  $k_{0,max} \simeq 100$ for the longest baselines, and $k_{0,max} \simeq 5$ for the shortest baselines. If the baseline distribution is such that the average baseline is half the maximum baseline, this strategy saves a factor of 2 in DFT operations and data rate.

\subsubsection{Efficient calculation of $X_{pq}[k]$ with sliding DFTs}
\label{sec:sliding}
We consider problem of computing DFT values for overlapping time windows. For typical array configurations and DMs, the number of samples in the DFT (N) is of order $10^5$, whereas the number of usable DFTs is of the order $k_{0, max} \simeq10^2$, meaning that computing a full FFT would result in a very large number of unused bins. In addition, we  do not require a DFT result every sample, which means a sliding window DFT result every $L<N$ samples is adequate.

A naive method to computing the sliding window DFT is to (1) compute the dot product of N input samples with a complex sinusoid of appropriate frequency, then (2) shift the input sequence by $L < N$ samples, and (3) compute the dot product on the shifted samples, with the same complex sinusoid. This naive method requires $N$ complex multiplications per $L$ samples, per DFT bin, and corresponds to an operations rate of roughly $f_s N/L$ per DFT bin.

\citet{Jacobsen03} describe a `the sliding DFT', a more efficient method for computing a small number of DFT bins in a sliding window manner. The sliding DFT is a recursive filter that produces a sliding window DFT output according to:

\begin{eqnarray}
S_k[n] = S_k[n -1] \exp {\left (- 2 pi  \pi j k/N \right )} - x[n - N] + x[N] \label{eq:sdft}
\end{eqnarray}

\noindent where $S_k[n]$ is the sliding window DFT output for sample $n$ and bin $k$, and $x[n]$ is the sampled input sequence. Equation \ref{eq:sdft} is effectively a moving average filter implemented as a Cascaded Integrator Comb (CIC), with a complex resonator embedded in the integrator feedback path. The sliding DFT has a operations rate of only $\simeq 3 f_s$ per DFT bin, which is significantly less than that required for the naive method.

In practice, we do not require an output every sample, so the operations rate can be further reduced by computing a block-based sliding DFT. In this case, we compute the partial DFTs, time-indexed by $m$ in blocks of $L$ samples:

\begin{equation}
V_k[m] = \sum_{n = 0}^{L-1}{\exp{\left ( - 2 \pi j  ( n + mL ) k /N \right ) x [ n + mL]}}
\end{equation}

\noindent and the and form the the DFT over the full number of samples $N$ by applying a moving average filter on the partial DFTs:

\begin{equation}
S_k[m] = S_k[m-1] + V_k[m] - V_k[m-N/L]. \label{eq:block_cic}
\end{equation}

\begin{figure}
\centering
\includegraphics[width=\linewidth]{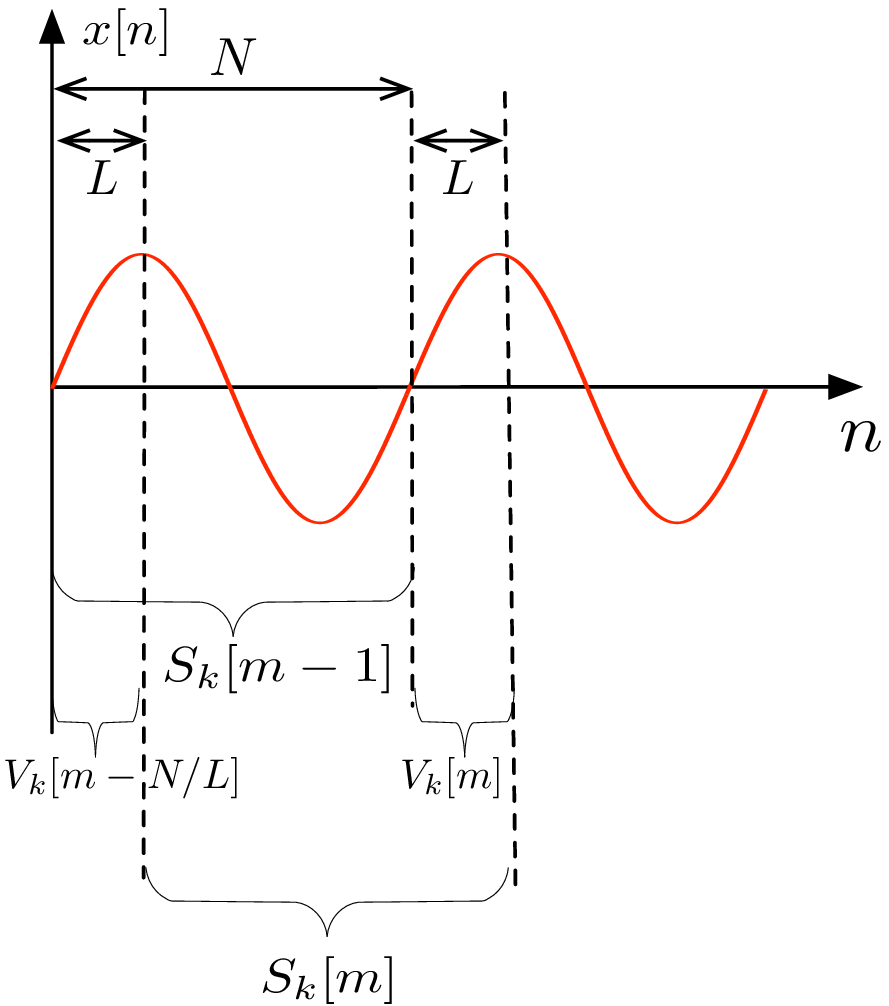}
\caption{The DFT can be efficiently computed in a sliding window manner (see Equation \ref{eq:block_cic}). For a DFT bin number $k$, the current value of the DFT bin ($S_k[m]$) is formed by taking the previous value of the DFT bin ($S_k[m - 1]$), adding the most recent partial DFT ($V_k[m]$) and subtracting the oldest partial DFT ($V_k[m-N/L]$).}
\label{fig:slidingwindows}
\end{figure}

This method is illustrated in Fig. \ref{fig:slidingwindows}.

The block-based sliding DFT has a lower operations rate than the sliding DFT, because the moving average (CIC) stage (Equation \ref{eq:block_cic}) operates at the block rate, rather than the sample rate, which resulting in an operations rate of  $\simeq f_s + 2 f_s/L = f_s(1 + 2/L) \simeq f_s$.

\subsubsection{Factorizing the DFTs}
\label{sec:factorisation}

\begin{figure}
\centering
\includegraphics[width= \linewidth]{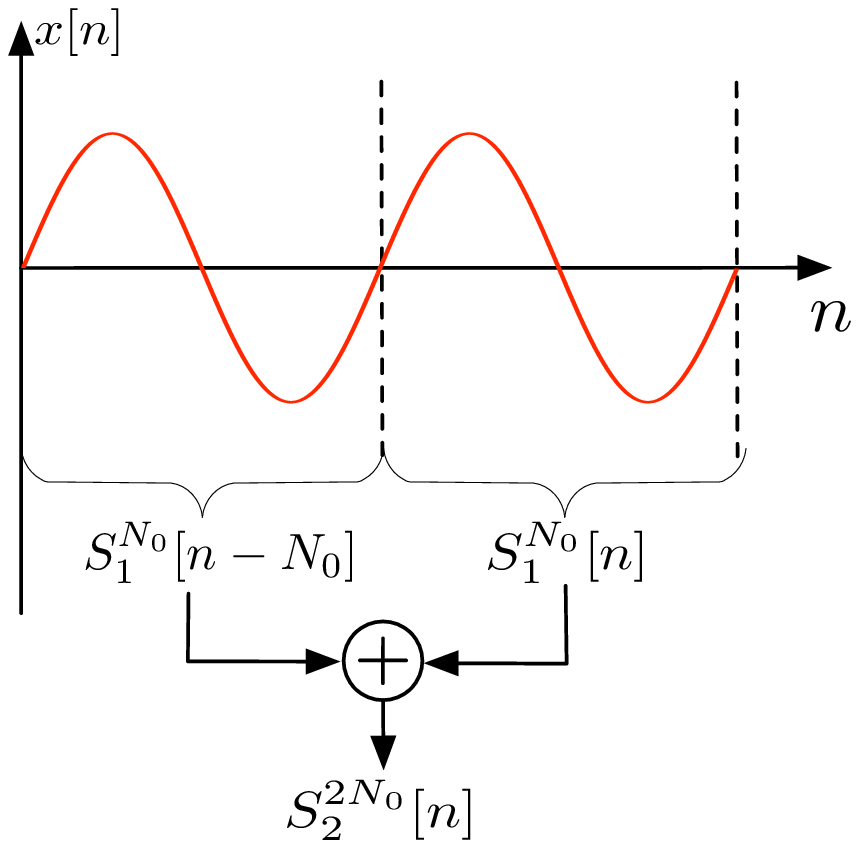}
\caption{The calculation of some DFTs can be factorised into the sum of two adjacent DFT results. In this example, we illustrate how to calculate the $k=2$  bin of the length $2N_0$ DFT ($S_2^{2 N_0}[n]$) by adding the results of two, adjacent, non-overlapping $k=1$ bins of the length $N_0$ DFT ($S_1^{N_0}[n] + S_1^{N_0}[n - N_0]$). In the notation of Section \ref{sec:factorisation}, this example corresponds to $k=d=D=2$.}
\label{fig:factorisation}
\end{figure}

As the DM and therefore the value of $\dot{f}$ is unknown, a search in $\dot{f}$ is required to maximize signal-to-noise ratio of pulse at unknown DM. This search through $\dot{f}$ is equivalent to varying the size of the DFT: $N$. One might choose to use a bank of DFTs, each with a length of $N= N_0 d$, where $N_0$ is the length of the DFT corresponding to the shortest DM of interest, and $d$ is a positive integer. If we have a bank of DFTs, each starting at the most recent sample and extending back in time by $N$ samples, then we can factorize some of the DFTs by noting that some of the basis functions for the long DFTs can be formed by concatenating the  basis functions for the short DFTs. By way of example, the result of the  $S_{2}$ bin for the length $2N_0$ window can be trivially computed by summing the adjacent, non-overlapping results of $S_{1}$  over the length $N_0$ window, as illustrated in Fig. \ref{fig:factorisation}.

If we write $S_k^N[n]$ as the DFT result for bin $k$ at sample time\footnote{In practice, one would compute the factorized DFTs on the sliding DFT block outputs indexed by $m$. We have kept the full sample rate $n$ here for clarity.} $n$ for a length $N$ DFT, we can say that a DFT of length $N_0 d$ can be computed from the sum of $D$ shorter length $N_0 d / D$ DFTs, if it can be written as:

\begin{equation}
S_{k}^{N_0 d}[n] = \sum_{d' = 0}^{D - 1}{S_{k/D}^{N_0 d / D}[n - d' N_0]}
\end{equation}

\noindent where $k$ and $d$ are integers.

The bin $S_{k}^{N_0 d}$ can be factorized if and only if $d/D$ and $k/D$ are integers, that is $d$ and $k$ must have a common, non-unity factor $D$ which implies that $d$ and $k$ cannot be co-prime. The probability of two random integers being co-prime is approximately 61 per cent \citep{HardyNumberTheory}, which implies that approximately 39 percent of DFT bins can be factorized. If the shorter DFT results are already available, computing the factorized DFT requires $D-1$ operations, which is trivial in comparison with $N$ operations to compute the full DFT. As a result, DFT factorization saves approximately 39 percent in complex operations.

\subsubsection{Efficiently computing negative DFT bins}
\label{sec:efficient_neg}

Computing the DFT output requires the multiplication of the complex input sample with the complex exponential. On a standard computer, the complex numbers are stored as real and imaginary parts, and the complex multiplication is performed in the following way:

\begin{eqnarray}
r_k [n] & = &x[n] \exp(-2 \pi j k n / N) \\
&= & (a + jb) (c - j d) \\
& = & ac + bd + j (-ad + bc) \label{eq:fpos}
\end{eqnarray}

\noindent where $r_f[k]$ is the result of the multiplication of the input sample with the sinusoid of frequency $k$,  $a$ and $b$ are the real and imaginary parts of the complex input sample, and $c$ and $d$ are the real and imaginary parts of the complex sinusoid.

Assuming the phase center is set to the center of the primary beam, the DFTs must be computed for frequencies over the range $[-k_{\rm 0, max}, k_{\rm 0, max}]$ to cover the full field of view. Therefore, each positive bin has a negative counterpart. To compute the negative frequency, we could also separately calculate:

\begin{eqnarray}
r_{-k} [n] & = &x[n] \exp(2 \pi j k n / N) \\
& = & (a + jb) (c + j d) \\
& = & ac - bd + j (ad + bc) \label{eq:fneg}
\end{eqnarray}

The calculation of $r_{k} [n] $ and $r_{-k} [n] $ naively requires 12 operations (8 multiplications and 4 additions). But the multiplications are common between the two results (Equations \ref{eq:fpos} and \ref{eq:fneg}), which means both results can be computed with only 8 operations (4 multiplications and 4 additions). This results in a saving of 33 percent over the naive implementation.

\subsection{Performance}

\subsubsection{Resolution}
From the definitions of Equations \ref{eq:pth} and \ref{eq:eth}, the amplitude of the response to a chirp on a given baseline is $|P(\theta)|^2 = D^2_N(x)$. Therefore we approximate the spatial resolution of the The Chirpolator as the Full Width Half Maximum (FWHM) of $D^2_N(x)$ on the longest baseline. The FWHM of $D^2_N(x)$ is defined by:

\begin{eqnarray}
D^2_N{(2 x_{\rm FWHM})} & = & \frac{1}{2} D^2(0) \\
\implies \frac{\sin(2 \pi x_{\rm FWHM})}{\sin (2 \pi  x_{\rm FWHM} / N)} & = & \frac{1}{\sqrt{2}} N
\end{eqnarray}

Taking the 3rd order Taylor expansion of the $\sin$ terms, and solving for the non-trivial solutions of $x$, we obtain:

\begin{eqnarray}
x_{\rm FWHM} & = & \frac{1}{\pi} \sqrt{ \frac{6 \left( 1 - 1/\sqrt{2} \right)}{1 - 1/N^2 \sqrt{2}} } \\
& \simeq & \frac{1}{\pi} \sqrt{6 \left (1 - 1/\sqrt{2}  \right )} \\
& \simeq & 0.844 \label{eq:fwhm}
\end{eqnarray}

To convert Equation \ref{eq:fwhm}, which is the width of the main lobe in units of DFT bins, to an angle, we rearrange Equation \ref{eq:f0Bbaseline}, which yields:

\begin{eqnarray}
\sin \theta & = & \frac{c k_{0} }{b_{pq} B} , \label{eq:sintheta}
\end{eqnarray}

\noindent and by applying the small angle formula, and substituting Equation \ref{eq:fwhm} as the DFT bin resolution (i.e. $\Delta k_0 = x_{\rm FWHM}$), we obtain the spatial resolution of The Chirpolator:
\begin{eqnarray}
\Delta \theta & = & \Delta k_0 \frac{c}{B b_{\rm max} } \\
& = &  0.844 \frac{c}{B b_{\rm max} } \label{eq:resolution}
\end{eqnarray}

\subsubsection{Operations rates}
\label{sec:chirp_compcost}

We compute the number of operations required to form  images of the full field of view of a telescope comprised of parabolic dishes. We keep to the convention of \citet{Cordes97} of counting complex operations, where a complex multiplication and accumulation is considered a single operation. As such, we do have not accounted for the 33 percent saving in floating point operations for the DFT as described in Section \ref{sec:efficient_neg}. Also, for clarity, we have not included the additional DFTs required to support nonlinear dispersion ($\delta_{\rm mix}$, see Section \ref{sec:nonlinear_dispersion}), as this substantially complicates the analysis, is only significant on the longest baselines and roughly balances the 33 percent saving described above.

To begin, we assume the half width beam of a parabolic reflector, at 25\% of the peak amplitude is \citep{Cordes97}:

\begin{equation}
\theta_{\rm max} = 0.585 \frac{\lambda}{D}.
\end{equation}

The full width of the beam at 25\% amplitude is $2 \theta_{\rm max}$ and we set the phase center to the center of the primary beam.  If we compute only the required DFT bins (as described in Section \ref{sec:only_dft_baselines}), the number of DFT bins that must be computed for a single DM over all baselines and the full primary beam, is given by:
\begin{eqnarray}
N_{\rm DFT}  & = & \sum_{p=0}^{M-1} \sum_{q=0, q \neq p}^{M-1} {  \frac{B}{c} b_{pq} 2 \sin \theta_{\rm max}} \\
& = & 2 \frac{B}{c}  \sin \theta_{\rm max}\sum_{p=0}^{M-1} \sum_{q=0, q \neq p}^{M-1} {b_{pq} } \\
& = &  2 \frac{B}{c} \sin \theta_{\rm max}  \frac{M}{2} (M - 1)  \bar{b} \\
& \simeq &  \frac{B}{c}   \theta_{\rm max}  M^2 b_{\rm max}
\end{eqnarray}

\noindent where we have employed the small angle formula for $\sin$, and assumed a distribution of baselines such that the average baseline length is approximately half of the maximum baseline length. Each of these DFT bins requires $\simeq f_s = B$ operations per second (using the block-based sliding DFT. See Section \ref{sec:sliding}), and assuming we measure $N_{\rm DM}$ dispersion measures then we require $N_{\rm DM}$ different DFT banks. The operations rate for the DFT step is, therefore, given by:

\begin{eqnarray}
\dot{C}_{\rm DFT} & = & B  N_{\rm DFT} N_{\rm DM} \rho_f  N_{\rm pol}\\
& \simeq & 0.585 \frac{\lambda}{D} \frac{B^2}{c}  M^2 b_{\rm max} N_{\rm DM}\rho_f N_{\rm pol}
\end{eqnarray}

\noindent where $\rho_f =0.61$ is a factor to account for factorizing the DFTs across the DM banks as described in Section \ref{sec:factorisation}, and $N_{\rm pol}$ is the number of polarizations.

To form an image for a given dispersion measure, a dot product with the truncated amplitude response function, across all baselines must be performed for every pixel. The number of pixels in an image is:

\begin{eqnarray}
N_{\rm pix} & = & \left ( \kappa_s \frac{ 2 \theta_{\rm max}}{\Delta \theta} \right )^2 \\
& = & 1.92 \left ( \kappa_s  \frac{  B b_{\rm max} \lambda}{c D} \right )^2
\end{eqnarray}

For each pixel, we require a dot product with the response function per baseline, implying

\begin{equation}
N_{\rm ops-per-pixel} = (2F + 1) M^2.
\end{equation}

An image is formed per DM bank at a rate given by $T_i \kappa_t$ where $T_i$ is the dispersion delay associated with the $i$th DM of interest, and  $\kappa_t \ge 1$ is the time over sampling factor. If we choose set of DM banks that is a geometric progression\footnote{We can choose a geometric progression for the DM bank lengths, and, when $N_i > N_0$, round $N_i$ to an integer multiple of $N_0/L$ to take maximum advantage from factorization as required in Section \ref{sec:factorisation}.} according to:

\begin{equation}
T_i = T_0 (1 + \epsilon) ^ i  \qquad 0 \le i < N_{\rm DM} \label{eq:Ti}
\end{equation}

\noindent with $\epsilon < 1$ an overlap factor which can be chosen by a trade-off between computation and SNR. The number of DM banks required to cover the range of DMs from $T_0$ to $T_{\rm max}$  is given by:

\begin{equation}
N_{\rm DM} \simeq \frac{ \log(T_{\rm max}/T_0) } {\log ( 1 + \epsilon )} \label{eq:ndms}
\end{equation}

\noindent and, images are produced at a rate:

\begin{eqnarray}
\dot{N}_{\rm image} & = & \sum_{i=0}^{N_{\rm DM} - 1} \frac{\kappa_t}{T_i} \\
& = &  \sum_{i=0}^{N_{\rm DM} - 1} \frac{\kappa_t }{T_0 (1 + \epsilon)^i} \\
& = & \frac{\kappa_t}{T_0} \sigma_{\rm DM}
\end{eqnarray}

\noindent where
\begin{equation}
\sigma_{\rm DM} = \frac{ 1 - (1 + \epsilon)^{- N_{\rm DM}+ 1 } }{ 1 - (1 + \epsilon)^{-1} }\label{eq:sigmadm}
\end{equation}

Finally, the operations rate for the imaging step is:

\begin{eqnarray*}
\dot{C}_{\rm img} & = & N_{\rm pix} N_{\rm ops-per-pixel} \dot{N}_{\rm image} N_{\rm pol} \\
& \simeq & 1.92 \left (  \kappa_s \frac {B b_{\rm max} \lambda}{c D} \right )^2 (2F + 1) M^2 \frac{\kappa_t}{T_0} \sigma_{\rm DM} N_{\rm pol}
\end{eqnarray*}

\subsubsection{Data rates}
\label{sec:chirp_iocost}
The DFT step takes in voltages for all antennas and produces $N_{\rm DFT}$ outputs sufficient to produce images at a rate of $\dot{N}_{\rm image}$. The data rate between the DFT step and the imaging step is therefore
\begin{equation}
R_{\rm DFT-out} = N_{\rm DFT}  \rho_f \dot{N}_{\rm image} N_{\rm pol} N_{\rm bytes-per-DFT-bin}. \label{eq:chirp_io}
\end{equation}

The data rate at the output of the imaging is:
\begin{equation}
R_{\rm img-out} =\dot{N}_{\rm image} N_{\rm pix} N_{\rm pol} N_{\rm bytes-per-pixel}.
\end{equation}

\section{The Chimageator: Analysis and implementation}
\label{sec:app_chimg}

In this appendix we describe methods for efficiently implementing The Chimageator. We also derive equations for the resolution and operations, and data rate requirements.

\subsection{Non-linear dispersion}
\label{sec:nonlinear_dispersion_chimg}

The Chimageator is also affected by non-linear dispersion. As with The Chirpolator (see Section \ref{sec:nonlinear_dispersion}), the mixing frequency in the non-linear case is no longer constant resulting non-linear trajectories in $n$-$k$ space and higher $k_{0, \rm max}$. The solutions may be to increase the time oversampling $\kappa_t$ and spatial oversampling $\kappa_s$ with additional phase correction to recover coherence.

As The Chimageator is not really very competitive in the near term (Fig. \ref{fig:askap_compcost}) we leave a detailed treatment for a future paper.

\subsection{Implementation optimizations}
\label{sec:chimg_optimizations}
\subsubsection{Optimizing operations in the imaging step}
\label{sec:chimg_optimising}

In order to form an image with The Chimageator, we need not sum across all possible trajectories. From Fig. \ref{fig:Chimageator}, one can see that there is a family of arrival angles and dispersions with the same gradient, but whose durations, $T$, differ. For trajectories on the same gradient, the results for all chirp durations can be computed by cumulative sum, i.e. the $P_{T_i}(\theta_i)$ can be computed recursively from the result $P_{T_{i-1}}(\theta_j)$ where $T_i$, $T_{i-1}$, $\theta_i$ and $\theta_j$ are chosen to have the same gradient in $n$-$k$ space. The gradient of the trajectory is given by:

\begin{equation}
\dot{k_0}(T, \theta) = \frac{1}{T} \frac{b_{\rm max}}{c} \sin(\theta) \label{eq:chimg_grad}
\end{equation}

The requirement for a shorter trajectory to be calculated from a longer trajectory implies:

\begin{eqnarray}
\dot{k_0} (T_i, \theta_i) &  \simeq &  \dot{k_0} (T_{i-1}, \theta_j) \\
\implies T_i    & \simeq & T_{i - 1} \frac{\sin \theta_i}{\sin \theta_j}
\end{eqnarray}

\noindent where the equivalence of the gradient can be traded depending on SNR and computational requirements.

Because of this recursive property, there are a much smaller number of independent calculations required to search through the DM (equivalent to the $T_i$) and $\theta$ space than one might naively expect. We require only enough operations to calculate the trajectories that end on the rectangle bounded by $N_{\rm max} = f_s T_{\rm max}$ in the $n$ axis, and $k_{0, max} = f_s b_{\rm max} \sin \theta_{\rm max}/c$ in the $k$ axis, as shown in Figure \ref{fig:Chimageator}. All shorter trajectories can be obtained as partial sums of calculation of the longer trajectory with the same (or similar) gradient.

This optimization works because all trajectories are linear, which means all the shorter trajectories can be constructed from partial sums of a single longer trajectory with the same gradient . When considering the non-linear dispersion, the trajectories are no longer linear and a short trajectory does not lie along the path of a single long trajectory. One possible solution is to consider the trajectories as piece-wise linear. The shorter trajectories can then be constructed from the piece-wise partial sums over a \emph{number} of long trajectories. We leave a detailed treatment of this approach to a future paper.

\subsubsection{Sampling}

For typical interferometers and dispersions, the gradient of the trajectory is reasonably small, which implies an integrate-and-dump operation after the gridding and FFT step can reduce the required data volumes and downstream processing requirements. If we assume a trajectory of duration $T$ is sampled $\kappa_t$ times, and assume the oversampling is proportional to the final DFT bin as follows:
\begin{eqnarray}
\kappa_t & = & k_{\rm 0, end} \kappa_{\rm t, 0} \\
& = & f_s \frac{b_{\rm max}}{c} \sin(\theta) \kappa_{\rm t, 0}
\end{eqnarray}

\noindent then we can produce a sequence of integrated samples indexed by $m$:

\begin{equation}
Y_k[m] = \sum_{n=0}^{f_s T / \kappa_t}{X_k[n + m f_s T / \kappa_t]}
\end{equation}

\noindent from which we can form the intensity image in Equation \ref{eq:chimg_pth} in much the same way, but at a reduced rate.

\begin{figure}
\centering
\includegraphics[width=\linewidth]{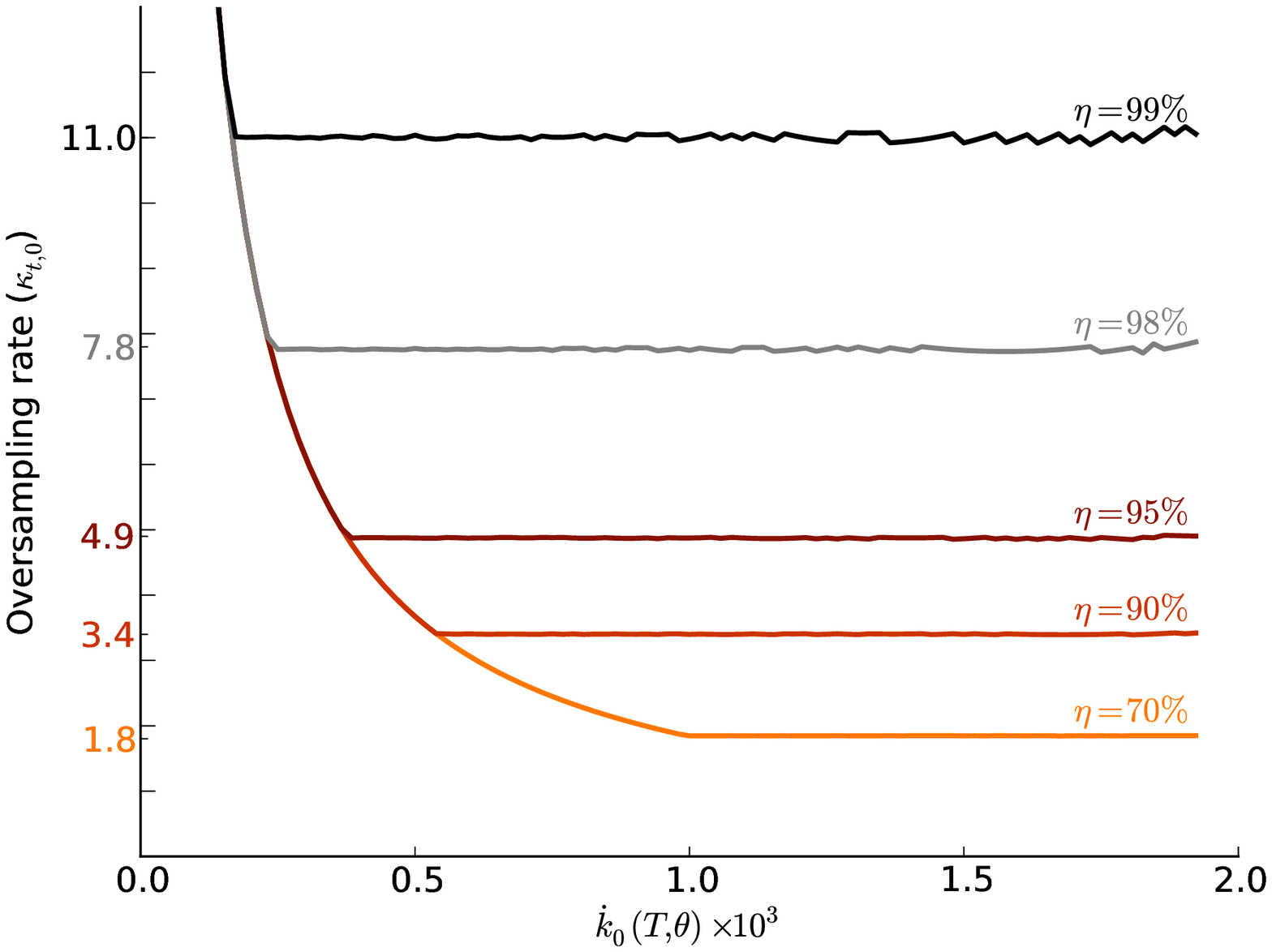}
\caption{The Chimageator can operate with non-linear dispersion as long as a sufficiently high oversampling rate is chosen. Above is the required oversampling factor ($\kappa_{\rm t, 0} = \kappa_t/k_{0, end}$) as a function of trajectory gradient ($\dot{k}$), to maintain a range of coherence loss levels. The simulated array had centre frequency 1.4~GHz, bandwidth 3~MHz, 4 antennas and 500~m spacing. The input signal was a linear chirp with dispersion delay corresponding to 20 cm$^{-3}$~pc. The trajectory gradient was calculated for 100 arrival angles from 0 to 3 degrees. The increase at small gradients is due to $k_{0, end} \simeq 0$ for small gradients but $\kappa_t=1$.
\label{fig:Chimageator_os_vs_grad}}
\end{figure}

A long integration time will smear out the signal and result in a loss of coherence, which results in a practical limit for how small $\kappa_{\rm t, 0}$ can be made. Figure \ref{fig:Chimageator_os_vs_grad} illustrates the amount of coherence loss which is achieved for a given value of $\kappa_{\rm t, 0}$. This plot suggests that oversampling the trajectory over 5 times during a its duration is sufficient to keep the coherence above 95 per cent.

\subsubsection{Spectral smearing}
As with The Chirpolator, when the instantaneous frequency of the chirp is between discrete DFT bins, the  energy is spread out along all DFT bins. To recover some SNR in this case, we sum along an $2F$ additional terms in the frequency direction (effectively widening the trajectory) to improve the SNR, in much the same way as described in Section \ref{sec:improving_snr}.

\subsection{Performance}
\subsubsection{Resolution}

The Chirpolator and Chimageator have the same resolution characteristics. This is demonstrated by considering $k_{0, max}$ for the two methods. With The Chimageator,  $k_{0, max}$ occurs when the sample number is the final sample of the chirp, i.e. $n = f_s T = B T$. For The Chirpolator, $k_{0, max}$ occurs on the the maximum baseline, $b_{max}$. In either case, it has a value:

\begin{equation}
k_{\rm 0, max} = B \frac{b_{\rm max}}{c} \sin(\theta_{\rm max})
\end{equation}

\noindent and the resolution is given by Equation \ref{eq:resolution}

\subsubsection{Operations rates}
\label{sec:chimg_compcost}

The first steps in Chimageator processing are the multiplication and gridding stages. Until now, we have assumed a uniform linear array, which makes gridding reasonably straightforward. For more complex geometries, a larger gridding kernel is required. A trade between the size of gridding support and the quality of the images is outside the scope of this paper, but for dimensioning purposes, one can consider a $7 \times 7$ pixel grid kernel, resulting in $N_{\rm ops-per-grid-point} \simeq 50$ including the multiplication of the voltages from the two antennas.

A grid point must be formed from each pair of antennas at the sampling rate, resulting in an operations rate for gridding of:
\begin{equation}
\dot{C}_{\rm gridding} = f_s \frac{M}{2}(M - 1) N_{\rm ops-per-grid-point}
\end{equation}

Assuming a spatial oversampling of $\kappa_s$ the number of pixels in the grid plane is:

\begin{equation}
N_{\rm pix} = (2 k_{\rm 0, max} \kappa_s) ^ 2
\end{equation}

\noindent The operations rate for the spatial FFT step is:

\begin{equation}
\dot{C}_{\rm FFT} = f_s N_{\rm pix} \log_2{N_{\rm pix}},
\end{equation}

\noindent and the operations rate for the integration step is:

\begin{equation}
\dot{C}_{\rm int} = f_s N_{\rm pix}.
\end{equation}

The total pre-integrator operations rate is therefore:

\begin{equation}
\dot{C}_{\rm pre-int-total} = \dot{C}_{\rm gridding} + \dot{C}_{\rm FFT}  + \dot{C}_{\rm int}
\end{equation}

The total pre-integrator operations rate is dominated by the FFT for sparse arrays, while for dense arrays, it is dominated by the gridding.

To compute the data and operations rates of imaging, we begin by assuming we use $N_{\rm DM}$  logarithmically spaced set of trial DMs as described in Equations \ref{eq:Ti} and \ref{eq:ndms}, and that each trajectory is sampled at the rate:

\begin{equation}
R_i = \frac{\kappa_t}{T_i}
\end{equation}

The computations are then broken into the two types of trajectory shown in Figure \ref{fig:Chimageator}: the trajectories with fixed angle $\theta_{\rm max}$ and variable $T_i$, and the trajectories with fixed time $T_{\rm max}$, and variable angle, $\theta_i$. In each case, a single trajectory requires $\kappa_t(2F + 1)$ operations per integration step. Therefore, the computation rate for a single trajectory is:

\begin{eqnarray}
\dot{C}_{\rm traj}  & =  & R_i \kappa_t(2F + 1) \\
& = & \frac{\kappa_t^2 (2F+ 1)}{T_i} \\
& = &  \frac{k_{\rm 0, end}^2 \kappa_{\rm t, 0}^2 (2F+ 1)}{T_i}
\end{eqnarray}

We now consider the total operations rate for the one-dimensional case and assuming half the beamwidth.

The fixed angle trajectories have fixed $k_{\rm 0,end} = k_{\rm 0, max}$, and variable $T_i$, resulting in an operations rate of:

\begin{eqnarray}
\dot{C}_{\theta_{\rm max}} & = & \sum_{i = 1}^{N_{\rm DM} }{\dot{C}_{\rm traj} } \\
& = &\sum_{i = 1}^{N_{\rm DM} }{ \frac{k_{\rm 0, end}^2 \kappa_{\rm t, 0}^2 (2F+ 1)}{T_i}} \\
& = & k^2_{\rm 0, max} \kappa^2_{\rm t, 0} (2F + 1) \sum_{i = 1}^{N_{\rm DM}}{\frac{1}{T_i}} \\
& = & k^2_{\rm 0, max} \kappa^2_{\rm t, 0} (2F + 1) \sigma_{\rm DM}
\end{eqnarray}

\noindent where $\sigma_{\rm DM}$ is defined in Equation \ref{eq:sigmadm}.

The fixed time trajectories, have fixed $T_i = T_{\rm max}$ and variable $k_{\rm 0,end}$, resulting in an operations rate of operations rate of:

\begin{eqnarray}
\dot{C} _{T_{\rm max}} & = & \sum_{k = 0}^{k_{\rm 0, max} \kappa_s}{\dot{C}_{\rm traj}} \\
& = &  \sum_{k = 0}^{k_{\rm 0, max} \kappa_s}{ \frac{k_{\rm 0, end}^2 \kappa_{\rm t, 0}^2 (2F+ 1)}{T_i}} \\
& = &  \sum_{k = 0}^{k_{\rm 0, max} \kappa_s}{\frac{(k/\kappa_s)^2 \kappa_{\rm t, 0}^2 (2F+1)}{T_i}} \\
& = & \frac{1}{T_{\rm max}} \frac{\kappa_{\rm t, 0}^2}{\kappa_s^2} (2F + 1)  \sum_{k = 0}^{k_{\rm 0, max} \kappa_s}{k^2} \\
& = & \frac{1}{6T_{\rm max}} \frac{\kappa_{\rm t, 0}^2}{\kappa_s} (2F + 1)k_{\rm 0, max}(\kappa_s k_{\rm 0, max} + 1)(2 \kappa_s k_{\rm 0, max} + 1) \\
& \simeq & \frac{1}{3 T_{\rm max}} \kappa_{\rm t, 0}^2 \kappa_s (2F + 1)k^3_{\rm 0, max}\end{eqnarray}

\noindent where $\kappa_s$ is the desired spatial oversampling. The total operations rate is the sum of the two sets in the one-dimensional case. In the two dimensional case, the computation is squared, so that the total operations rate in 2-dimensions, for the full beamwidth is:

\begin{eqnarray}
\dot{C}_{\rm Total} = \left ( 2 ( \dot{C}_{\theta_{\rm max}} + \dot{C} _{T_{\rm max}})\right)^2 \label{eq:chimg_ctotal}
\end{eqnarray}

\subsubsection{Data rates}
\label{sec:chimg_iocost}
Assuming this integrate-and-dump operates at the highest rate $R_0 = \kappa_t/T_0$ and the longer integrations can be formed from the short integrations in the imaging step, then the data rate at the output of the integrate and dump step is:

\begin{eqnarray}
R & = & N_{\rm pix}N_{\rm bytes-per-pix} R_0\\
& = & (2 k_{\rm 0, max} \kappa_s)^2 N_{\rm bytes-per-pix}  k_{\rm 0, max} \kappa_{\rm t, 0}/T_0 \\
& = & 4 \kappa_s^2  \kappa_{\rm t, 0} k^3_{\rm 0, max} N_{\rm bytes-per-pix} / T_0 \label{eq:chimg_io}
\end{eqnarray}

\end{document}